\documentclass[aps,prb,twocolumn,showpacs]{revtex4}
\usepackage{bm}
\usepackage{graphicx}

\begin{document}
\title{On Equilibrium Dynamics of Spin-Glass Systems}

\author{A. Crisanti}
\email{andrea.crisanti@roma1.infn.it}

\author{L. Leuzzi}
\email{luca.leuzzi@roma1.infn.it}
\affiliation{Dipartimento di Fisica, Universit\`a di Roma ``La Sapienza''}
\affiliation{Istituto Nazionale Fisica della Materia, Unit\`a di Roma, 
             and SMC,
             P.le Aldo Moro 2, I-00185 Roma, Italy}
\affiliation{Istituto Studi della Complessit\'a (ISC), CNR, 
             Via dei Taurini 19, I-00185 Roma, Italy}

\begin{abstract}
We present a critical analysis of the Sompolinsky theory of equilibrium 
dynamics.
By using the spherical $2+p$ spin glass model
we test the asymptotic static limit of the Sompolinsky solution
showing that it fails to
yield a thermodynamically stable solution.
We then present an alternative formulation, based on the
Crisanti, H\"orner and Sommers [Z. f\"ur Physik {\bf 92}, 257 (1993)]
dynamical solution of the spherical $p$-spin spin glass model,
reproducing a stable static limit that coincides, in the case
of a one step Replica Symmetry Breaking Ansatz, with the solution at the 
dynamic free energy
threshold at which the relaxing system gets stuck off-equilibrium.
We formally extend our analysis to any number of Replica Symmetry 
Breakings $R$.
In the limit $R\to\infty$ both formulations lead to the Parisi 
anti-parabolic differential equation. This is the special case, though,
where no dynamic blocking threshold occurs.
The new formulation does not contain the additional order 
parameter $\Delta$ of the Sompolinsky theory.

\end{abstract} 


\maketitle

\section{Introduction}
In the last years a great deal of work has been devoted to the study 
of the so called {\it off-equilibrium} dynamics of glassy systems, i.e., the
dynamics on time scales large enough to discard the initial condition but
not to ensure equilibrium. This large amount of work has left aside the
analysis of the equilibrium dynamics, i.e., 
the dynamics which should lead to the static properties derived from 
statistical mechanics.

When discussing the equilibrium dynamics of spin-glass systems, one usually
refers to the Sompolinsky solution.\cite{Somp81b}
Sompolinsky assumed that the relaxation dynamics of a spin glass system
occurs via a set of large relaxation times $t_x$, all of which become infinite
in the thermodynamic limit, reflecting the
hierarchical order of free-energy barriers or states of the spin glass phase.
By incorporating explicitly this assumption into the relaxation dynamics
of the Sherrington-Kirkpatrick (SK) model\cite{SheKir75,SheKir78} he
was able to construct a consistent mean-field dynamical theory
that, in the limit of an infinite series of relaxation times, is described by 
two continuous order parameters functions: the overlap function $q(x)$, 
measuring the 
amount of correlation that has not yet decayed, and $\Delta(x)$,   
representing the {\it anomalous} contribution to the response function.
As in the static calculation the variable $x$ can be defined to vary in the 
interval $[0,1]$ with $x=1$ corresponding to the shortest (though 
infinite) time 
scale and $x=0$ to the longest. With this definition $q(x)$ is a 
nondecreasing function while $\Delta(x)$ is a non-increasing 
function with  boundary condition $\Delta(1)=0$.

In the static limit the Sompolinsky solution, in general, does not
coincide with the Parisi static solution of the Full Replica Symmetry
Breaking (FRSB) phase.\cite{Parisi80} De Dominicis, Gabay and
Orland\cite{deDomGabDup82} (DGO) have, indeed, shown that the static
limit of the Sompolinsky solution can be derived from a static
calculation with replicas by using a replica symmetry breaking (RSB)
scheme different from Parisi's one.  The two schemes, however, coincide in
the so called {\it Parisi Gauge}, i.e. choosing the function
$\Delta(x)$ such that $d\Delta(x)/dx = - x\,dq(x)/dx$.\cite{Somp81b}

In this work we reconsider the Sompolinsky solution, we show the
instability of its static limit and we check the validity of an
alternative solution, originally proposed by Crisanti, Horner and
Sommers (CHS),\cite{CriHorSom93} in the context of a generic R Replica
Symmetry Breaking scenario.

\begin{figure}
\includegraphics[width=.49\textwidth]{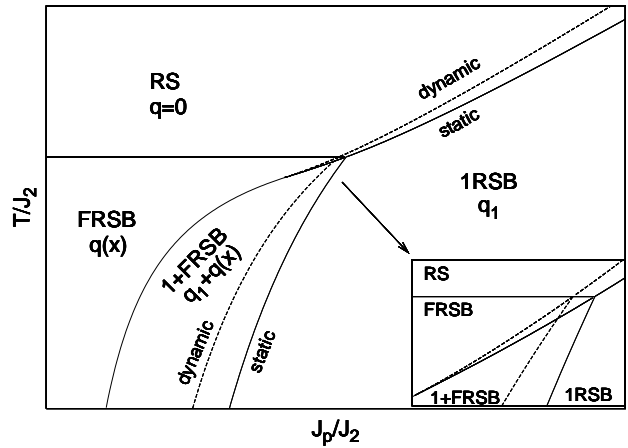}
\label{fig:static_pd}
\caption{Qualitative sketch of the $T-J_p$ phase diagram of the $2+p$
spin spherical model with $p>3$ in the mean field approximation.  Four
phases are displayed.  (i) RS/paramagnetic for which the overlap order
parameter is zero; (ii) 1RSB/structural glass-like, where the order
parameter is the single overlap $q_1$; (iii) FRSB/spin glass whose
order parameter is a continuous function; (iv) 1+FRSB with an order
parameter consisting in a function $q(x)$ plus a single number $q_1$
representing the self-overlap and such that $q_1>q(1)$.  The full
curves are the static transition lines, whereas the dotted ones are
the dynamic ones. Notice that dynamic transitions are different from
static ones only when an apart step of RSB occurs. Indeed this happens
for the RS/1RSB transition, for which the dynamic transition line is
rederived in the proper dynamic contest in Sec.  \ref{ss:CHS-Stab}, as
well as for the 1RSB/1-FRSB and FRSB/1-FRSB transitions (in the latter
case only in a small region, see inset).  For $p=3$ only the RS and
the 1RSB phases are present.  }.
\end{figure}

Our testing bench is the $2+p$-spin interacting spherical model whose
static properties have been studied by the authors in previous works.
\cite{CriLeu05,CriLeu06} Such model, for $p>3$, displays a rich phase
diagram that we show in Fig. \ref{fig:static_pd}.  It contains a
Replica Symmetric (RS) phase (i.e. a phase in which the RS Ansatz
yields a thermodynamically stable solution), a one step Replica
Symmetry Breaking (1RSB) phase, an infinite steps RSB phase and even a
phase consisting of an infinite, continuous (or {\em full}), set of
RSBs plus an apart step of RSB (we call it 1-FRSB solution).  Besides
this, it has the further advantage, with respect to e.g. the SK or the
Ising $p$-spin models, of being exactly solvable in each one of the
phases.  In the same model it is, therefore, possible to analytically
check the validity of the Sompolinsky solution (and any alternative
proposal) both in a phase where the thermodynamics is known to be 1RSB
and in one where it is FRSB.  The relaxation dynamics for the present
model is illustrated in Sec.  \ref{sec:dynmod}. There we also solve
the equations of motion making use of two simple Ansatz (the dynamic
analogues of the RS  Ansatz and of the Sommers Ansatz,
respectively).  This should help to fix notation and concepts and
serves as a starting point for the subsequent discussion.

The line of investigation  proceeds, then,  along the
following steps.
\begin{description}

\item{Sompolinsky solution. Sec. \ref{sec:SompSol}. } \\ Reconsidering
in detail the derivation of the Sompolinsky solution, we observe that,
in a Parisi 1RSB-stable phase, it tends to a static solution different
from the one of Parisi as the time goes to infinity.  This is not
dramatic, since there is no reason preventing the dynamic limit from
being different from the thermodynamic solution (corresponding to the
global minimum of the free energy landscape of the system). Indeed,
for 1RSB systems, it is a well known property that in a quenching
procedure from high temperature the dynamics gets stuck at a {\em
threshold} free energy level strictly above the equilibrium
one.\cite{KirThi87}

\item{Stability of the Sompolinsky static limit.
Sec. \ref{sec:StabSomp}. } \\ Always working in the 1RSB phase of the
$2+p$ spherical model and using the DGO formalism, we check the
stability of the Sompolinsky solution in its static limit.  We find
that it is thermodynamically unstable (details of the proof of the
instability are reported in Appendix \ref{app:StabSom}).  We further
generalize this result to the case of a dynamics described by any
finite number $R$ of diverging relaxation times.\cite{Note-Luca}
Eventually, we analyze the $R\to \infty$ limit in which the
Sompolinsky static limit and the static Parisi solution
coincide, provided one fixes the Parisi's gauge, and we address
the reasons of the qualitative difference with the behavior at finite
$R$.

\item{CHS Solution, static limit and
stability. Sec. \ref{sec:CHS-R}. }  \\ We propose an alternative
formulation of the equilibrium dynamics of spin glass systems, based
on the CHS dynamical solution of the spherical $p$-spin
model.\cite{CriHorSom93} This is a solution apparently similar to the
Sompolinsky's one, but based on slightly different assumptions, that,
however, turn out to be crucial in curing the instability of the
latter.  As well as Sompolinsky's the CHS solution tends, as $t\to
\infty$, to a solution different from the Parisi one.  The explicit
computation of the solution on the 1RSB-stable phase of the $2+p$
spherical model shows that the infinite time limit coincides with the
corresponding Parisi solution at the threshold free energy and that,
unlike Sompolinsky's, it is marginally stable in that limit.  The same
formalism is effective for any number of steps and the limit $R\to
\infty$ is considered as well.  Details are reported in Appendix
\ref{app:Dyn2p}.
\end{description}

Finally, in the Appendices A and B, 
we report the DGO derivation of the Sompolinsky 
solution and discuss its connection with the Parisi solution in the 
FRSB phase.

\section{The dynamical model}
\label{sec:dynmod}
To illustrate the equilibrium dynamics of spin glass systems we use 
the spherical $2+p$ spin model defined by the Hamiltonian
\begin{equation}
\label{eq:ham}
H =  \frac{r}{2}\sum_i\sigma_i^2
     - \sum_{i<j}^{1,N}J^{(2)}_{ij}\sigma_i\sigma_j
           - \sum_{i_1<\ldots <i_p}^{1,N}J^{(p)}_{i_1\ldots i_p}
           \sigma_{i_1}\cdots\sigma_{i_p}
\end{equation}
where $p$ is an integer equal or larger than $3$ and $\sigma_i$
are $N$ continuous real spin variables which range from $-\infty$ to 
$+\infty$ subject to the global spherical constraint
\begin{equation}
\label{eq:sphconst}
\sum_{i=1}^{N} \sigma_i^2 = N.
\end{equation}
The coupling strengths $J^{(s)}_{i_1\ldots i_s}$ ($s=2,3,\ldots$) 
are quenched independent identically distributed
 Gaussian variables of variance
\begin{equation}
\label{varVp}
  \overline{\left(J^{(s)}_{i_1 i_2..i_s}\right)^2} = 
   \frac{s!\, J_s^2}{2\,N^{s-1}}, 
  \qquad i_1 < \cdots < i_s.
\end{equation}
and mean zero.
The scaling with the system size $N$ ensures an extensive free energy and hence
a well defined thermodynamic limit $N\to\infty$. Without 
loosing in generality one may take either $J_2$ or $J_p$ equal to $1$
since this only amounts to a rescaling of the temperature $T$.
Finally, the parameter $r$ is a Lagrange multiplier needed to impose the
spherical constraint. In the following, when discussing the FRSB and 
1-FRSB phases of the model, we implicitly assume $p>3$. 

The relaxation dynamics of the model is described by the Langevin equation
\begin{equation}
\Gamma_0^{-1}\,\partial_t\,\sigma_i(t) = 
               \frac{\delta\,\beta H}{\delta \sigma_i(t)} + 
	       \xi_i(t)
\end{equation} 
where $\beta^{-1} = T$ is the temperature, $\Gamma_0^{-1}$ a microscopic 
time-scale and the noise $\xi_i(t)$ a Gaussian variable of zero mean 
and variance
\begin{equation}
\langle \xi_i(t)\, \xi_j(t')\rangle_{\xi} = 
            2\,\Gamma_0^{-1}\,\delta_{ij}\,\delta(t-t')
\end{equation}
which ensures the proper equilibrium distribution.

In dynamical calculations the quantities of interest are product of
spins averaged over the thermal noise and disorder. Of particular interest
are the local spin correlation function
\begin{equation}
   C(t,t')= \overline{\langle\sigma_i(t)\,\sigma_i(t')\rangle_{\xi}}
\end{equation}
and the average local response function
\begin{equation}
G(t,t')= \frac{\delta\,\overline{\langle\sigma_i(t)\rangle_{\xi}}}
              {\delta\, \beta h_i(t')}, 
 \qquad t\geq t'
\end{equation}
where $h_i(t)$ is an external magnetic field.\cite{Note:beta}

Using the Martin-Siggia-Rose formalism \cite{MSR73} in the
path integral formulation\cite{DeDomPel78,BauJanWag76} the correlation and
response functions can be obtained from a generating functional for dynamic
correlations and response functions. The disordered average can be done 
directly on the generating functional without using replicas since the
generating functional is normalized to one.\cite{DeDom78}
The calculation is now rather standard and we do not report it but
give directly the results.
The interested reader can find more details in 
Refs. [\onlinecite{SomZip82,CriSomp87,CriHorSom93}].

In the thermodynamic limit $N\to\infty$ the dynamics reduces to a
single-spin self-consistent non-Markovian dynamics described by the 
equation:
\begin{equation}
\Gamma_0^{-1}\,\partial_t\sigma(t) = -\beta r \sigma(t) +
               \int_{t_0}^{t}\, dt'\, \Sigma(t,t')\,\sigma(t') + \eta(t)
\end{equation}
where $t_0$ is some initial time
and $\eta(t)$ a Gaussian noise 
with zero mean and variance
\begin{equation}
\langle\eta(t)\,\eta(t')\rangle = 2\,\Gamma_0^{-1}\,\delta(t-t') 
                          + \Lambda(t,t').
\end{equation}
The vertex $\Lambda$ and the self-energy $\Sigma$ of the $2+p$ model
are given by
\begin{equation}
\label{eq:Lambda}
\Lambda(t,t') \equiv \Lambda[C(t,t')] 
              = \mu_2\, C(t,t') + \mu_p\, C(t,t')^{p-1}
\end{equation}
\begin{eqnarray}
\label{eq:Self}
\Sigma(t,t') &\equiv& \Lambda'[C(t,t')]\,G(t,t') \nonumber\\
              &=& \left[\mu_2 + \mu_p\, (p-1)\, C(t,t')^{p-2}\right]\, G(t,t')
\end{eqnarray}
where
\begin{equation}
\mu_2 = (\beta J_2)^2,\ \mu_p = \frac{p}{2} (\beta J_p)^2
\end{equation}
and $\Lambda'(x) \equiv d\Lambda(x)/dx$.

The correlation and response functions must be evaluated self-consistently from
the single-spin dynamics as
\begin{equation}
C(t,t') = \langle \sigma(t)\, \sigma(t')\rangle, \qquad
G(t,t') = \frac{\partial \langle\sigma(t)\rangle}{\partial \beta h(t')}
\end{equation}
where the average $\langle(\cdots)\rangle$ is over the random noise $\eta(t)$.

Since we are interested in the equilibrium correlation and response
function we take the initial time $t_0$ equal to $-\infty$ so that
two-times quantities become function of the time difference only (in
other words we are in time translational invariant
regime).\cite{BauJanWag76} To work in Fourier space we introduce the
transformed functions
\begin{eqnarray}
C(\omega) &=& \int_{-\infty}^{+\infty}\, dt\, {\rm e}^{i\omega t}\,C(t), \\
G(\omega) &=& \int_{0}^{+\infty}\, dt\, {\rm e}^{i\omega t}\,G(t)
\end{eqnarray}
The single-spin equation of motion 
then reads
\begin{equation}
\label{eq:eqm}
\sigma(\omega) = G(\omega)\, \eta(\omega)
\end{equation}
where $G(\omega)$ obeys the Dyson equation
\begin{eqnarray}
\label{eq:Dyson}
G^{-1}(\omega) &=& \beta r - i\Gamma_0^{-1}\omega - \Sigma(\omega) \nonumber\\
               &=& G_0^{-1}(\omega) - \Sigma(\omega)
\end{eqnarray}
and $\eta(\omega)$ is a Gaussian variable of zero mean and variance
\begin{equation}
\label{eq:eta}
\langle\eta(\omega)\,\eta(\omega')\rangle = 2\pi\,\delta(\omega+\omega')\,
   \left[2\Gamma_0^{-1} + \Lambda(\omega)\right]
\end{equation}
In the Fourier space the correlation function $C(\omega)$ is given by
\begin{equation}
\label{eq:Com}
C(\omega) = \langle\sigma(\omega)\,\sigma(-\omega)\rangle,
\end{equation}
where the average is over the noise $\eta(\omega)$. 
The Fluctuation Dissipation Theorem (FDT)\cite{Note:beta}
\begin{equation}
G(t) = -\theta(t)\,\partial_t\,C(t)
\end{equation}
is recast, in Fourier space, as
\begin{equation}
\label{eq:FDTo}
C(\omega) = \frac{2}{\omega}\,{\rm Im}\, G(\omega)
\end{equation}
The FDT implies that 
the static susceptibility $G(\omega=0)$ reads
\begin{equation}
G(\omega=0) = C(t=0) - C(t\to\infty)
\end{equation}
This reduces to $G(\omega=0) = 1$ when
 the spherical constraint is imposed ($C(t=0)=1$) and the decay  to zero
of $C(t)$ for large times is assumed.


\subsection{The ``replica symmetric'' solution}

Before introducing the Sompolinsky solution we consider the derivation
of the static limit of the dynamics assuming the existence of a time
persistent contribution to the correlation function. We will
eventually see [Eqs. (\ref{eq:Lambdars},\ref{eq:Gammars})] that, in
the limit $\omega \to 0$ (or $t\to\infty$), it leads to a static
solution equivalent to a replica symmetric (RS) one.

The strategy for constructing the solution in the spin glass phase is the
following. First one assumes that in the spin glass phase the correlation 
function $C(t)$ decays to a finite value
\begin{equation}
\lim_{t\to\infty} C(t) = q > 0.
\end{equation}
The parameter $q$ is called the Edwards-Anderson order parameter and 
represents the time-persistent part of the correlation.
This implies that $C(\omega)$ is of the form
\begin{equation}
\label{eq:Crs}
C(\omega) = \tilde{C}(\omega) + 2\pi\,q\,\delta(\omega)
\end{equation}
where 
$\tilde{C}(t)= C(t)-q$ is the finite-time part of $C(t)$, decaying to 
zero as $t\to\infty$.

For the spherical model the single-spin equation of motion 
(\ref{eq:eqm}) is linear and the self-consistent equation for $q$ can be 
easily derived just substituting the equation of motion 
(\ref{eq:eqm}) into the definition (\ref{eq:Com}) of $C(\omega)$ 
and extracting the time-persistent part.
However, we derive it in the following in a more general 
 way.

Inserting Eq. (\ref{eq:Crs}) for the correlation function 
into the definition of the vertex function $\Lambda(\omega)$ one has
\begin{eqnarray}
\label{eq:vert}
\Lambda(\omega) &=& \int_{-\infty}^{+\infty}\, dt\, 
                     {\rm e}^{i\omega t}\,\Lambda(t) \nonumber \\
                &=& \int_{-\infty}^{+\infty}\, dt\, 
                     {\rm e}^{i\omega t}\,
                   \bigl[\Lambda[C(t)] - \Lambda(q) + \Lambda(q)\bigr]
\nonumber \\
                &=& \tilde{\Lambda}(\omega) + 2\pi\,\Lambda(q)\,\delta(\omega)
\end{eqnarray}
where $\tilde{\Lambda}(\omega)$ contains only contributions from the 
finite-time part of the correlation function and is, hence,
  non-singular for  $\omega\to 0$. 
Starting from this separation, and looking at Eq. (\ref{eq:eta}), 
the noise $\eta(\omega)$ can be 
split into the sum of two independent Gaussian noises
\begin{equation}
\label{eq:noise-split}
\eta(\omega) = \phi(\omega) + z(\omega)
\end{equation}
where $\phi(\omega)$ is
defined by the finite-time part of the vertex function:
\begin{equation}
\langle\phi(\omega)\,\phi(\omega')\rangle_{\phi} = 
   \,2\pi\,\delta(\omega+\omega')\,
   \left[2\Gamma_0^{-1} + \tilde{\Lambda}(\omega)\right]
\end{equation}
while $z(\omega)$ by the time-persistent part
\begin{equation}
\label{eq:noisez}
\langle z(\omega)\, z(\omega')\rangle_z = 2\pi\,\delta(\omega+\omega')\,
                   \Lambda(q)\,2\pi\,\delta(\omega).
\end{equation}
The two noises $\phi$ and $z$ represent, hence, respectively, the 
the ``{\it fast}'' and ``{\it slow}'' parts of the noise $\eta$.

By definition, $q$ is the remaining part of the correlation function
 once that the correlations induced by the fast  part
of the noise have died out. 
As a consequence the self-consistent equation 
for $q$ reads
\begin{equation}
\label{eq:qrs}
q = \left\langle \langle\sigma\rangle_{\phi}^2\right\rangle_{z}
\end{equation}
where $\langle\sigma\rangle_{\phi}=\langle\sigma(\omega=0)\rangle_{\phi}$ 
is the static average value of $\sigma(\omega)$ induced by the noise 
$\phi(\omega)$ in presence of a fixed static random noise $z$.

To solve the equation of motion and find  
$\langle\sigma\rangle_{\phi}$ a relation between $G(\omega)$ and
$C(\omega)$ is needed. Assuming that, as in ordinary phase transitions, the
effect of an external perturbation will die out on finite-time scales,  
the full response function $G$ is related to the finite-time part
$\tilde{C}$ of $C$ by the FDT
\begin{equation}
\label{eq:FDT0}
\tilde{C}(\omega) = \frac{2}{\omega}\,{\rm Im}\, G(\omega)
\end{equation}
which, in turn, implies  
\begin{equation}
\langle\phi(\omega)\,\phi(-\omega)\rangle_{\phi} = 
       -\frac{2}{\omega}\,{\rm Im}\, G^{-1}(\omega).
\end{equation}
This relation ensures that the noise $\phi$ acts as a {\it thermal} noise 
and hence
$\langle\sigma\rangle_{\phi}$ is the magnetization induced in
thermal equilibrium by the static Gaussian field $z$,
\begin{eqnarray}
\langle\sigma\rangle_{\phi} &=& \bar m(z) \nonumber\\
&=&  \frac{
    \int_{-\infty}^{+\infty} d\sigma\,\sigma\,
          \exp{[-\frac{1}{2}G^{-1}(\omega=0)\,\sigma^2 + z\sigma]}
      }
      {
    \int_{-\infty}^{+\infty} d\sigma\,
          \exp{[-\frac{1}{2}G^{-1}(\omega=0)\,\sigma^2 + z\sigma]}
      } \nonumber\\
     &=& G(\omega=0)\, z
\label{eq:mz}
\end{eqnarray}
Since
the equation of motion of the spherical $2+p$ spin glass model is linear in 
$\sigma(\omega)$, this result can be also obtained by averaging directly the
equation of motion (\ref{eq:eqm}) over the noise $\phi(\omega)$ and
taking the limit $\omega\to 0$.

Inserting this expression into Eq. (\ref{eq:qrs}) and using 
Eq. (\ref{eq:noisez}) one ends up with
\begin{equation}
q = G(\omega=0)^2\,\Lambda(q).
\end{equation}
Eventually, the expression of the static susceptibility can be readily obtained
with the help of the FDT relation (\ref{eq:FDT0}) and reads
\begin{equation}
G(\omega=0) = \tilde{C}(t=0) - \tilde{C}(t\to\infty) = 1-q.
\end{equation}
We, then, end up with the following self-consistent equation for $q$
\begin{equation}
\label{eq:Lambdars}
\Lambda(q) = \frac{q}{(1-q)^2}
\end{equation}
that coincides with the static RS solution of the spherical $2+p$ spin glass
model.\cite{CriLeu06}

The dynamical stability of this solution requires that the $\omega\to 0$ limit
of the kinetic coefficient, or generalized damping function, 
$\Gamma(\omega)$, must be non-negative. Its inverse is defined as 
\begin{eqnarray}
\label{eq:damp}
\Gamma^{-1}(\omega) =
 i\frac{\partial G^{-1}(\omega)}{\partial\omega}
              = \Gamma_0^{-1} - i\frac{\partial}{\partial\omega}\Sigma(\omega)
\end{eqnarray}

Inserting the form (\ref{eq:Crs}) of the correlation function 
into the definition of the self-energy $\Sigma(\omega)$,
and using manipulations similar to those used for extracting the singular part 
of $\Lambda(\omega)$, we have
\begin{equation}
\label{eq:Srs}
\Sigma(\omega) = \tilde{\Sigma}(\omega) + \Lambda'(q)\,G(\omega).
\end{equation}
In the limit $\omega\to 0$ one then obtains\cite{Note:Gamma}
\begin{eqnarray}
\Gamma^{-1}(\omega=0) &=& \lim_{\omega\to 0} \Gamma^{-1}(\omega)
=i\lim_{\omega \to 0}\frac{G^{-1}(-\omega)-G^{-1}(\omega)}{2  \omega}
\nonumber
\\
&=&
    \frac{
      \Gamma_0^{-1} + \frac{\partial}{\partial\omega}
                   {\rm Im}\, \tilde{\Sigma}(\omega = 0)
         }
         {
      1 - \Lambda'(q)\,G(\omega=0)^2
         }.
\end{eqnarray}
The numerator describes the decay of the finite-time part and is,
thus, positive. The requirement $\Gamma(\omega=0)\geq 0$ leads, then,
to the condition
\begin{equation}
\label{eq:Gammars}
1 - \Lambda'(q)\,G^2(\omega=0) = 
      1 - \Lambda'(q)\,(1-q)^2 \geq 0
\end{equation}
In terms of the static replica formalism, the above expression exactly
  coincides with the De Almeida-Thouless\cite{AlmeThou78} stability
  condition derived from the stability analysis of the RS saddle
  point. \cite{CriLeu06}

The replica symmetric solution is unstable everywhere in the low temperature
phase, thus this solution is correct only in the paramagnetic phase up to
the critical point where $\Gamma(\omega=0) = 0$.
Below this point a new 
solution is needed. 


\subsection{The {\em Sommers} Solution}
\label{ss:SomSol}
In the previous derivation of the  solution the FDT was assumed to 
hold between the full response function $G$ and the finite-time 
part $\tilde{C}$ of the correlation function. 
The failure of the replica symmetric solution to 
describe the relaxation in the spin-glass phase suggests that in this phase
the presence of a time-persistent part in the correlation function must 
reflect itself also in the response to an external perturbation 
with an anomalous contribution to the response function. This extra
contribution occurs, however, only in the static susceptibility, 
i.e., {\it exactly} at zero frequency:
\begin{equation}
\label{eq:Gsom}
G(\omega) = \tilde{G}(\omega) + \Delta\,\delta_{\omega,0}
\end{equation}
where $\delta_{\omega,0}$ is the Kronecker delta and $\Delta$
is the discontinuity between the static susceptibility and the 
$\omega\to 0$ limit of the dynamic susceptibility $G(\omega)$:
\begin{equation}
\Delta = G(\omega=0) - \lim_{\omega\to 0} G(\omega).
\end{equation}
The static limit of this solution is known as Sommers
solution.\cite{Somm78,Somm79} As before, the non singular finite-time
$\tilde{G}$ of the response function is related to the finite-time
part $\tilde{C}$ of the correlation function by the FDT
\begin{equation}
\label{eq:FDT1}
\tilde{C}(\omega) = \frac{2}{\omega}\,{\rm Im}\, \tilde{G}(\omega).
\end{equation}
Inserting the expressions (\ref{eq:Crs}) and (\ref{eq:Gsom}) for the 
correlation and response functions into the Dyson 
equation (\ref{eq:Dyson}), and making use of Eqs. (\ref{eq:noise-split}) and 
(\ref{eq:Srs}), 
a straightforward algebra leads to the following equation of motion  
\begin{equation}
\label{eq:eqmS}
\sigma(\omega) = \tilde{G}(\omega)\,\left[\phi(\omega) + H(\omega)\right]
\end{equation}
where $H(\omega)\equiv H(z)$ is the {\it effective} static noise 
\begin{eqnarray}
H(\omega) &=&   z(\omega) 
              + \Lambda'(q)\,\Delta\,\delta_{\omega,0}\,\sigma(\omega)
\nonumber\\
          &=&   z(\omega)
              + \Lambda'(q)\,\Delta\,\delta_{\omega,0}\,
                \langle\sigma\rangle_{\phi}
\end{eqnarray}
In the second expression we used the fact that, because of the
Kronecker delta $\delta_{\omega,0}$, only the part of $\sigma(\omega)$
which is nonzero at $\omega=0$ may contribute to the static field
$H(z)$.  This part is the static magnetization $\bar m(z) =
\langle\sigma\rangle_{\phi}$ induced by the static noise $z$.  The
product $\delta_{\omega,0}\,\langle\sigma\rangle_{\phi}$ is, however,
ill-defined since it contains the product of the functions
$\delta(\omega)$ and $\delta_{\omega,0}$, both having vanishing width.
To give a meaning to this product one introduces a finite-width
representation of the Dirac and Kronecker delta functions: 
$\epsilon$:
\begin{equation}
\lim_{\epsilon\to 0} \delta_{\epsilon}(\omega) = \delta(\omega), \quad
\lim_{\epsilon\to 0} \Delta_{\epsilon}(\omega) = \delta_{\omega,0}
\end{equation}
Then $\delta_{\omega,0}\,\langle\sigma\rangle_{\phi}$ is defined as  the 
$\epsilon\to 0$ limit of the convolution of $\delta_{\epsilon}(\omega)$ and 
$\Delta_{\epsilon}(\omega)$.

If the width of $\Delta_{\epsilon}(\omega)$ is much smaller than the one of
$\delta_{\epsilon}(\omega)$ then the contribution of 
$\langle\sigma\rangle_{\phi}$ to the static field $H(z)$ 
is vanishing, and one gets back the $\Delta=0$ solution. 
In the opposite limit,\cite{Somp81b}
$\Delta_{\epsilon}(\omega)\,\delta_{\epsilon}(\omega)\to \delta(\omega)$ and
the full magnetization $\bar m(z) = \langle\sigma\rangle_{\phi}$ contributes to
the static field $H(z)$.

The finite-time parts $\tilde{G}$ and $\tilde{C}$ of the response and 
correlation functions are related by the FDT. As a consequence 
the fast noise $\phi$ is a {\it thermal} noise and, hence,  
$\langle\sigma\rangle_{\phi}$ is the static thermal equilibrium 
magnetization induced by the static field $H(z)$ [as it was in
Eq. (\ref{eq:mz})]:
\begin{eqnarray}
\bar m(z) &=& \tilde{G}(\omega=0)\,H(z) \nonumber\\
                            &=& \tilde{G}(\omega=0)\,
      \left[\sqrt{\Lambda(q)}\,z + \Delta\,\Lambda'(q)\,\bar m(z)\right]
\label{eq:mSom}
\end{eqnarray}
where we have rescaled the Gaussian noise $z$ to have 
$\langle z^2\rangle_z = 1$.
Solving for $\bar m(z)$ 
and inserting the result into Eq. (\ref{eq:qrs}) we obtain 
the self-consistent equation
\begin{equation}
\label{eq:qSom}
q = \left[\frac{1-q}
               {1 - \Delta\,(1-q)\,\Lambda'(q)}
    \right]^2\, \Lambda(q)
\end{equation}
where we used the relation $\tilde{G}(\omega=0) = \tilde{C}(t=0) = 1 -q$
following from FDT.

The equation for the anomalous term $\Delta$ 
is obtained from the 
definition of the static susceptibility and reads:
\begin{equation}
\left\langle
                   \left.\frac{\partial \bar m(z)}
{\partial \beta h}\right|_{h=0}
             \right\rangle_{z}
=1-q+\Delta 
\end{equation}
where $h$ is a static external field. Adding the field $h$ to the 
equation of motion (\ref{eq:eqmS}) and inserting the resulting $\bar 
m(z)$ into the
above equation we have, after some algebra, 
\begin{equation}
\label{eq:DeltaSom}
 \frac{1-q}{1 - \Delta\,(1-q)\,\Lambda'(q)} = 1-q+\Delta 
\end{equation}
The two self-consistency 
equations (\ref{eq:qSom}, \ref{eq:DeltaSom})
 can be rewritten in the form
\begin{equation}
\Lambda(q) = \frac{q}{\left(1-q+\Delta\right)^2}
\end{equation}
\begin{equation}
\label{eq:GammaSom}
\Lambda'(q) = \frac{1}{(1-q)(1-q+\Delta)}
\end{equation}
from which one readily sees that for $\Delta=0$ the solution reduces
to the replica symmetric solution at criticality: $\Gamma(\omega=0)=0$
[cfr.  Eqs. (\ref{eq:Lambdars}) and (\ref{eq:Gammars})].  The
dynamical stability requirement on the damping function $\Gamma(\omega
=0) \geq 0$ for the Sommers solution is still given by the
Eq. (\ref{eq:Gammars}). As a consequence, for any $\Delta>0$ the
solution has a positive $\Gamma^{-1}(\omega = 0)$.  This apparently
hints that this solution (or, at least, its static limit) is
physically stable.  However, as noted by Hertz, \cite{Her83} exactly
at $\omega = 0$ the time persistent part of the response function
[Eq. (\ref{eq:Gsom})] should not change the relative static solution,
thus implying a negative non-linear susceptibility.  We will
reconsider this point when we will present the study of the static
limit of the present dynamics in the DGO formalism in Sec.
\ref{sec:StabSomp}


\section{The Sompolinsky Solution}
\label{sec:SompSol}
The Sommers solution assumes that there are only two relevant
time-scales, a short time-scale related to the finite-time part of the
motion, and a long time-scale -- actually infinite in the
thermodynamic limit -- related to the time-persistent part of the
motion.  This scenario is clearly too limitative for the description
of the spin-glass phase where different time-scales are involved.

The Sompolinsky solution extends the Sommers solution to the case of
many different long times-scales, all of which diverge in the
thermodynamic limit. To be more specific one assumes that there are
$R$ different relaxation times $t_r$, $r=1,\ldots,R$. As $N\to\infty$
all times go to infinity with the prescription $t_r/t_s\to\infty$ if
$r<s$. The short-time scale relaxation time, which can be identified
with $t_{R+1}$, is proportional to $\Gamma_0^{-1}$ and remains finite
for $N\to\infty$.

\begin{figure}
\includegraphics[scale=1.0]{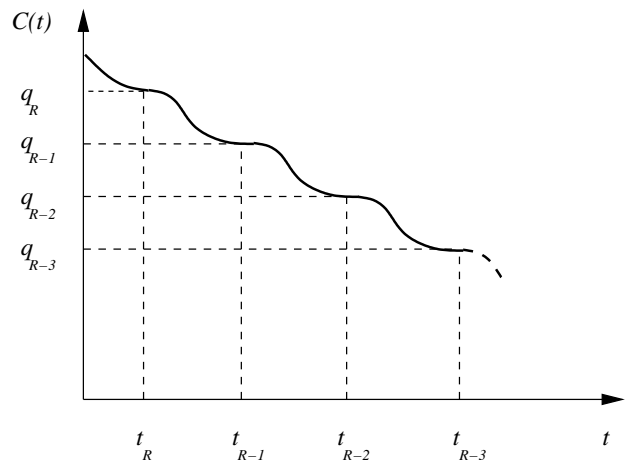}
\caption{Schematic form of correlation function with many relaxation 
         time-scales.
        }
\label{fig:corr}
\end{figure}

In each time interval, or sector, $t_{r+1}\ll t\ll t_r$.  The
 relaxation process with characteristic times less than $t_r$ have
 already relaxed to equilibrium, while those with longer (or equal)
 relaxation times have not relaxed yet.  For each time interval
 $t_{r+1}\ll t\ll t_r$ we can introduce an order parameter
\begin{equation}
\label{eq:qr}
q_r = {\rm T}_r\,\lim_{t\to\infty} C(t), \qquad r=0,\ldots, R
\end{equation}
where the ``time ordered limit'' ${\rm T}_r\,\lim_{t\to\infty}$ is defined as
\begin{equation}
{\rm T}_r\,\lim_{t\to\infty} := \lim_{t_R\to\infty}\cdots
                    \lim_{t_{r+1}\to\infty}\lim_{t\to\infty}\lim_{t_r\to\infty}
		    \cdots\lim_{t_0\to\infty}
\end{equation}
The overlap $q_r$ measures the time-persistent part
 of the correlation function in the 
interval $[t_{r+1},t_r]$, see Fig. \ref{fig:corr}. 
With this definition $q_R$ coincides with the Edwards-Anderson order
parameter previously defined.  Moreover, we have introduced the
additional level $r=0$ associated with the longest time-scale, i.e.,
the equilibration time-scale, of the model. The overlap $q_0$
represents, then, the asymptotic equilibrium value (that is equal to
zero in absence of an external magnetic field).

The next step is to split away from the full correlation and response
functions the time-persistent parts as for the Sommers solution. The
functions $C$ and $G$, thus, are still of the form (\ref{eq:Crs}) and
(\ref{eq:Gsom}) with finite-time parts related by the FDT
(\ref{eq:FDT1}), but now\cite{Somp81b}
\begin{equation}
\label{eq:qSomp}
q\,\delta(\omega) \to \sum_{r=0}^{R}\, (q_r - q_{r-1})\, 
                \delta_{\epsilon_r}(\omega)
\end{equation}
and 
\begin{equation}
\label{eq:DeltaSomp}
\Delta\,\delta_{\omega,0} \to -\sum_{r=0}^{R}\, \dot{\Delta}_r\, 
                                 \Delta_{\epsilon_r}(\omega)
\end{equation}
where $\dot{\Delta}_r$ is the anomalous contribution to the response functions,
$\delta_{\epsilon}(\omega)$ and $\Delta_{\epsilon}(\omega)$ are finite-width
representations of the Dirac and Kronecker delta functions,
and $\epsilon_r = 1/t_r$. Here and in the following we use the convention
that all quantities of negative sub-index are zero.

As for the Sommers solution, the noise $\eta(\omega)$ is decomposed
into the sum of a fast ({\it thermal}) noise $\phi$ and a slow noise
$z$ composed by the sum of independent {\it slow} noises $z_r$ of zero
mean and variance
\begin{equation}
\langle z_r(\omega)\,z_r(-\omega)\rangle_r = 
            \left[\Lambda(q_r) -\Lambda(q_{r-1})\right]\,
            \delta_{\epsilon_r}(\omega).
\end{equation}
From the definition of $q_r$ and the fact that 
the noise $z_r$ acts as a static noise only up to the time-scale
$t_r$, the order parameter $q_r$ is, thus, given by
\begin{equation}
\label{eq:qzr}
q_r = \langle \bar m_r^2\rangle_z, \qquad r = 0,\ldots, R
\end{equation}
where $\bar m_r\equiv \bar m_r(\{z\})$ is the static magnetization
induced at scale $r$ by the slow noise $z$ and the average
$\langle\cdots\rangle_z$ is over all static noises $z_r$.  Clearly
$\bar m_r$ is a function of $z_0,\ldots,z_r$ only since all others
noises $z_s$ with $s>r$ have died out.  The magnetization $\bar m_r$
can be obtained from the magnetization $\bar m_R$ induced by the noise $z$
on the shortest time scale by integrating out the noises $z_s$ with
$s=r+1,\ldots,R$:
\begin{equation}
\label{eq:mzr}
\bar m_r = \int\,\prod_{s=r+1}^{R}\, Dz_s\, \bar m_R(z_0,\ldots,z_R)
\end{equation}
where $Dz_s\equiv P(z_s)\,dz_s$ and $P(z_s)$ is the probability
distribution of $z_s$.

One then proceeds as in Sec. \ref{sec:dynmod}, inserting
Eqs. (\ref{eq:Crs},\ref{eq:Gsom}, \ref{eq:qSomp}, \ref{eq:DeltaSomp})
for $C$ and $G$ and Eq. (\ref{eq:noise-split}) for the noise
$\eta(\omega)$ into the equation of motion and looking at its static
limit in order to derive the equations for the thermal equilibrium
magnetizations.  As we have seen in the study of the Sommers solution,
in this limit one has to deal with the products
$\delta_{\epsilon_r}(\omega)\,\Delta_{\epsilon_r}(\omega)$.  Clearly
one has
\begin{eqnarray}
\delta_{\epsilon_r}(\omega)\,\Delta_{\epsilon_s}(\omega)\simeq 
  &\delta_{\epsilon_r}(\omega)& \qquad \mbox{if $r<s$}
\\
\delta_{\epsilon_r}(\omega)\,\Delta_{\epsilon_s}(\omega)= &0&
\qquad \mbox{if $r>s$}
\end{eqnarray}
since $\epsilon_r/\epsilon_s\ll 1$ for $r<s$. 
{\em Yet, for
$s=r$ the product is ill-defined.} 

Sompolinsky solves the problem with the assumption that for 
$\epsilon\ll 1$ the width of the function $\delta_{\epsilon}(\omega)$ is  
much smaller than the width of $\Delta_{\epsilon}(\omega)$,\cite{Somp81b}
and the product goes like
\begin{equation}
\label{eq:SompAss}
\delta_{\epsilon}(\omega) \,  \Delta_{\epsilon}(\omega) \simeq 
        \delta_{\epsilon}(\omega), \qquad \epsilon\ll 1.
\end{equation}
With this assumption each level $r$ contributes
to the effective field $H(z)$ with the full magnetization
$\bar m_r$. As a consequence the self-consistent equation for
$\bar m_R$ reads [cfr. Eq(\ref{eq:mSom})]
\begin{equation}
\label{eq:mRSomp}
\bar m_R\{z\} = (1-q_R)\,H(\{z\})
\end{equation}
\begin{equation}
\label{eq:HSomp}
H(\{z\}) = \sum_{r=0}^{R}\, 
      \left[\sqrt{\Delta_r}\,z_r - \Delta'_r\,\bar m_r(\{z\})\right]
\end{equation}
where we used the identity $\tilde{G}(\omega=0) = 1-q_R$, we
rescaled the Gaussian variables $z_r$ in order to have
$\langle z_r^2\rangle_r = 1$  and we defined
\begin{equation}
\Delta_r = \Lambda(q_r) - \Lambda(q_{r-1}), \quad\quad
\Delta'_r = \Lambda'(q_r)\, \dot{\Delta}_r
\end{equation}
We anticipate  that the cause of 
instability of the static limit of the Sompolinsky solution (studied in Sec.
\ref{sec:StabSomp}), 
is hidden right in the conjecture expressed by Eq. (\ref{eq:SompAss}).
We will come back
to this problem and we will show how to overcome it in Sec. \ref{sec:CHS-R},
 where we will analyze the CHS solution.

The equation for the anomalous term $\dot\Delta_r$ 
follows directly from its definition: 
$\dot\Delta_r$ represents the anomalous contribution on scale $r$ 
to the static susceptibility.
The total anomalous contribution to the static susceptibility 
from the short-time scale up to scale $r$ is, then,
\begin{equation}
\int\prod_{s\leq r}\,Dz_s\, 
   \left.\frac{\partial \bar m_r}{\partial \beta h_r}\right|_{h_r=0}
=1-q_R - \sum_{s=r}^R\,\dot\Delta_s 
\end{equation}
where $h_r$ is a static external field active up to the temporal 
scale labeled by $r$, so that 
$\partial \bar m_r / \partial h_s = 0 $ if $r<s$.
The presence of $h_r$ just adds the term $\beta h_r$ to the $r$ contribution
to $H(z)$,
in Eq. (\ref{eq:HSomp}), and this implies
\begin{eqnarray}
\label{eq:DSomp}
\int\prod_{s\leq r}\,Dz_s\, 
   \frac{1}{\sqrt{\Delta_r}}\frac{\partial \bar m_r}{\partial z_r} 
&=&\frac{1}{\sqrt{\Delta_r}}
\left\langle\frac{\partial \bar m_r}{\partial z_r}\right\rangle_{z}
\\
\nonumber 
&=&1-q_R - \sum_{s=r}^R\,\dot\Delta_s
\end{eqnarray}
since $\bar m_r$  only  depends on $z_0,\ldots,z_r$.

Eqs. (\ref{eq:qzr}, \ref{eq:mzr}, \ref{eq:HSomp})
and (\ref{eq:DSomp}) together with the expression (\ref{eq:mRSomp}) for
$\bar m_R$
constitute the Sompolinsky solution for the spherical $2+p$ spin glass model. 
The Sommers solution is recovered by taking $R=0$.


\subsection{Sompolinsky's functional and  explicit solution of the $2+p$ 
spherical model}
The Sompolinsky solution can be obtained from the Sompolinsky 
functional,\cite{Somp81b} that, for the spherical $2+p$ spin glass model,
reads
\begin{eqnarray}
\label{eq:fsSomp}
-\beta f_{\rm S} &=& -\beta f_0(q_R) 
  + \frac{1}{2}\sum_{r=0}^R\,q_r\,\Delta'_r 
\\
\nonumber
&\phantom{=}&
  + \int\prod_{r=0}^R Dz_r
   \left[\frac{1}{2}\sum_{r=0}^R \Delta'_r\,\bar m_r^2(\{z\}) 
+ \phi(H(\{z\}))\right]
\end{eqnarray}
where 
\begin{equation}
\phi(H) = \frac{1}{2} (1 - q_R)\,H^2
\end{equation}
and
\begin{eqnarray}
-\beta f_0(q_R) &=& -\frac{1}{2}\Bigl[g(q_R) + \Lambda(q_R) (1-q_R) 
\nonumber\\
&\phantom{]}&
                -\frac{1}{1-q_R} -\log(1-q_R)\Bigr] 
 -\beta f_{\infty}.
\end{eqnarray}
The function $g$ is such that 
\begin{equation}
\frac{d g(q)}{dq} = \Lambda(q)
\end{equation}  
and the term $f_{\infty}$ is the infinite temperature limit of the
free energy, whose explicit form is only needed for computing
thermodynamic quantities.  The Sompolinsky equations follows from
stationarity of $f_{\rm S}$ with respect to variations of $\bar m_r$,
$\dot\Delta_{r}$ and $\Delta_r = \Lambda(q_r) - \Lambda(q_{r-1})$.

For the spherical $2+p$ spin glass model 
equation (\ref{eq:mzr}) for the local magnetization $\bar m_r$
can be explicitly solved. 
After a simple algebra one gets
\begin{equation}
\label{eq:mzSomp}
\bar m_r = \bar m_{r-1} + \frac{1-q_R}{1+(1-q_R)\,\sum_{s=r}^{R}\Delta'_s}\,
                \sqrt{\Delta_r}\,z_r
\end{equation}
that, with Eq. (\ref{eq:qzr}), leads to the following equations 
for the order parameter $q_r$ with $r= 1,\ldots,R$
\begin{equation}
\label{eq:qrSomp}
q_r - q_{r-1} = \left[\frac{1-q_R}
                      {1+(1-q_R)\,\sum_{s=r}^{R}\Delta'_s}\right]^2\,
               \Delta_r
\end{equation}
and $q_0$
\begin{equation}
\label{eq:q0Somp}
q_0 = \left[\frac{1-q_R}{1+(1-q_R)\,\sum_{s=r0}^{R}\Delta'_s}\right]^2
      \left[\Lambda(q_0) - (\beta\,h)^2\right]
\end{equation}
In the last equation we have added an external field $h$ to make $q_0$ finite.
Finally from Eq. (\ref{eq:DSomp}) we have
\begin{equation}
\label{eq:DDSomp}
1-q_R - \sum_{s=r}^R\,\dot\Delta_s = 
         \frac{1-q_R}{1+(1-q_R)\,\sum_{s=r}^{R}\Delta'_s}.
\end{equation}


\subsection{Comparison between the Parisi solution and the the static limit of 
the Sompolinsky solution}
The static solution for the spherical $2+p$ spin glass model within the Parisi 
$R$-RSB scheme, as obtained in Ref. [\onlinecite{CriLeu06}], consists of the 
following self-consistency equations:
\begin{eqnarray}
\label{eq:qrPar}
\Lambda(q_r) - \Lambda(q_{r-1}) &=& \frac{q_r - q_{r-1}}
       {\chi_r\, \chi_{r+1}}, \quad
r = 1,\ldots, R
\\
\label{eq:q0Par}
\Lambda(q_0) &=& \frac{q_0}{\chi_0^2}
	     - (\beta\,h)^2
\\
\label{eq:DDPar}
\Lambda'(q_r) &=& \frac{1}{\chi_{r+1}^2}
\end{eqnarray}
where
\begin{equation}
\label{eq:chiP}
    \chi_r = 1-q_R + \sum_{s=r}^R\,m_s(q_s-q_{s-1}),  \quad 
r = 0,\ldots, R
\end{equation}
The quantities $0<m_r<1$ are the RSBs parameters,
i.e., the sizes of the blocks of the Parisi $R$-RSB scheme
in the continuation from integer to real numbers in the
limit $n\to 0$, where $n$ is the total number of replicas.

The equations yielding the infinite time limit of the 
Sompolinsky solution for the spherical $2+p$ spin glass model, i.e. Eqs.
(\ref{eq:qrSomp}, \ref{eq:q0Somp}, \ref{eq:DDSomp}),
can be written in the equivalent form
\begin{eqnarray}
\label{eq:qrSomp1}
\Lambda(q_r) - \Lambda(q_{r-1}) &=& \frac{q_r - q_{r-1}}{\chi_r^2}, \quad 
r = 1,\ldots, R
\\
\label{eq:q0Somp1}
\Lambda(q_0) &=& \frac{q_0}{\chi_0^2}
	     - (\beta\,h)^2
\\
\label{eq:DDSomp1}
\Lambda'(q_r) &=& \frac{1}
  {\chi_r\,\chi_{r+1}}
\end{eqnarray}
where
\begin{equation}
\label{eq:chiSomp}
    \chi_r = 1-q_R - \sum_{s=r}^R\,\dot\Delta_s,  \quad 
r = 0,\ldots, R
\end{equation}

A simple inspection of the two sets of equations reveals that the
Sompolinsky solution cannot be reduced to the Parisi solution, not
even fixing the so called Parisi gauge $\dot\Delta_r =
-m_r(q_r-q_{r-1})$.  This implies that, for any finite value of $R$,
the Sompolinsky solution differs from the Parisi solution.

When the number of time sectors (or RSBs in the static counterpart) is
sent to infinite, however, the static limit of Sompolinsky solution
can be formally reduced to the Parisi solution,\cite{Somp81b} provided
that the gauge $d\Delta(x) = -x\, dq(x)$ is set.  The functions
$d\Delta(x)=\dot\Delta(x)\,dx$ and $q(x)$ are the limit functions of
$\dot\Delta_r$ and $q_r$ as $R\to\infty$. The parameter $x\in[0,1]$ is
the continuous limit of the series $\{m_r\}$.

This is more easily seen by using the replica 
derivation of the Sompolinsky solution introduced by 
DGO.\cite{deDomGabOrl81}
By using this approach it can be shown, see Appendix \ref{app:DGO}, 
that the Sompolinsky solution
can be derived from stationarity with respect to $q_r$ and $\dot\Delta_r$ 
of the DGO functional
\begin{eqnarray}
\label{eq:fDGOr}
-2\beta f_{\rm DGO}^{(R)} &=& -g(q_R) 
                - \sum_{r=0}^{R}\,\Lambda(q_r)\dot\Delta_r 
                + (\beta\,h)^2\chi_0 \nonumber\\
	&\phantom{=}&	+ \log(1-q_R) 
                + \sum_{r=0}^R\frac{q_r-q_{r-1}}{\chi_r}
\end{eqnarray}
with $\chi_r$ is given by Eq. (\ref{eq:chiSomp}).
We have neglected the term $f_{\infty}$, irrelevant to our discussion.

In the limit $R\to\infty$ in the FRSB phase we have $q_r-q_{r-1}\to dq$ while
$\dot\Delta_r\to (d\Delta(q)/dq)\,dq = \dot\Delta(q)\,dq$. 
As a consequence, the sums can be replaced by integrals and the DGO functional
for the spherical $2+p$ model becomes
\begin{eqnarray}
-2\beta f_{\rm DGO}^{\infty} &=& 
                - \int_{0}^{1}\,dq\,\Lambda(q)\dot\Delta(q) 
                - (\beta\,h)^2\int_{0}^{1}\,dq\,\dot\Delta(q) \nonumber\\
	&\phantom{=}&	+ \ln(1-q(1)) 
                - \int_{0}^{q(1)}\,\frac{dq}
                                    {\int_{q}^{1}\,dq'\,\dot\Delta(q')}
\label{eq:fDGOi}
\end{eqnarray}
where $q(1)=\lim_{R\to\infty}q_R$, and we have extended the definition
of $\Delta(q)$ to the whole interval $[0,1]$ defining
\begin{equation}
  \begin{array}{lcl}
 \dot\Delta(q) = 0  && \mbox{if}\ 0\leq q <  q_0 \\
&& \\
 \dot\Delta(q) = -1 && \mbox{if}\ q(1) < q\leq 1
\end{array}
\end{equation}
to have a more compact expression.

The analogous calculation can be performed within the Parisi scheme.
When the stable phase of the $2+p$ model is yielded by a FRSB solution, the
 Parisi 
functional is
\begin{eqnarray}
-2\beta f_{\rm P}^{(\infty)} &=&
            \int_{0}^{1} dq~ x(q)\, \Lambda(q) 
            + (\beta\,h)^2\int_{0}^{1}\,dq\,x(q) \nonumber\\
&\phantom{=}&
            + \ln\left(1 - q(1)\right)
            + \int_{0}^{q(1)} \frac{dq}{\int_{q}^{1} dq'\, x(q')}
\label{eq:rpl-f}
\end{eqnarray}
where $x(q)$ is the inverse function of $q(x)$. It is easy to see that 
the two functionals, Eqs. (\ref{eq:fDGOi}) and (\ref{eq:rpl-f}),
coincide in the Parisi gauge $\dot\Delta(q) = -x(q)$.

The fact that the two solutions differ might not be a problem.
Indeed, systems whose thermodynamics is described by a 1RSB stable
phase display the well known property of having a dynamic solution -- at
which the system relaxation gets arrested -- that is different from the
static solution.  This arrest is due to the presence of very many
metastable states of infinite lifetime lying at a free energy level
higher than the one of the global minima, selected, instead, by the static
solution.

The apparent paradox of having different solutions at finite $R$ but
coinciding ones for $R\to\infty$ can be solved by inspecting the
$R\to\infty$ limit of the DGO and Parisi RSB schemes.  It can be shown
that for any finite $R$ the difference between the Parisi and the
DGO-Sompolinsky theories is at least of order $O((q_r-q_{r-1})^2)$,
which is finite for finite $R$ but vanishes as $R\to\infty$.  The
gauge invariance of the Parisi equation for the order parameter $q(x)$
which follows from the DGO-Sompolinsky theory just reflects the
reparametrization invariance of the Parisi equation due to the
arbitrary definition of the variable $x$ in the Parisi scheme.  The
Parisi Gauge $d\Delta(q)/dq = -x(q)$ is the definition of the function
$x(q)$ whose $q$ derivative is right the probability density of overlaps.


\section{Stability in the replica formalism}
\label{sec:StabSomp}

The results just described rise the question on the validity of the
Sompolinsky solution. Is it a different but yet acceptable solution?
This question is better answered considering the phase of the
spherical $2+p$ spin glass model where the 1RSB Parisi Ansatz is known
to be stable.\cite{CriLeu06}
 In the 1RSB phase there is only one long time-scale, so that the
appropriate dynamical solution should be given by the Sompolinsky
solution with $R=0$, i.e., the Sommers solution of Sec.
\ref{ss:SomSol}. There, we have shown that, in the dynamic limit for infinite
times (i.e. $\omega\to 0$), the equation of motion appeared to be well
defined.  Here, we inspect more thoroughly the Sommers solution {\em
exactly} at $\omega = 0$ and we show that it turns out to be {\it
unstable} everywhere in the 1RSB phase.

The present analysis is carried out in the DGO formalism
(cfr. Appendix \ref{app:DGO}) that, for $R=0$, is analogous to the
Parisi RSB formalism with $R=1$.
Our approach is a straightforward generalization of the procedure
adopted in Ref. [\onlinecite{CriSom92}] to study the
stability of the  1RSB solution {\`a la Parisi} in
the spherical $p$-spin model.

The replicated free energy density as a function of the overlap matrix
${\bm q}$ reads:
\begin{eqnarray}
\label{eq:free2}
-\beta f[{\bm q}] &=& \frac{1}{2n} \sum_{ab}^{1,n} g(q_{ab})
 +\frac{1}{2n}\ln\det {\bm q} + s(\infty)
\\
\label{eq:free3}
g(x) &=& \frac{\mu_2}{2} x^2 + \frac{\mu_p}{p} x^p.
\end{eqnarray}
where $\mu_p = (\beta J_p)^2 p / 2$ and $ 
s(\infty)= (1+\ln 2\pi) / 2$ is the entropy per spin at infinite 
temperature. The parameter $n$ is the number of replicas.

The elements of the symmetric $n\times n$ real matrix ${\bm q}$ are
\begin{equation}
\label{eq:qab}
q_{ab} = \frac{1}{N}\sum_{i=1}^N\, \sigma_i^a\,\sigma_i^{b},
\quad\quad a, b = 1, \ldots, n.
\end{equation}
The spherical constraint, Eq. (\ref{eq:sphconst}), implies that 
the diagonal elements of the matrix $\bm{q}$  are all equal to one:
$q_{\alpha\alpha} = \overline{q} = 1$.

 The saddle point 
equation reads, in the $n\to 0$ limit,
\begin{equation}
\label{eq:spab}
\Lambda(q_{\alpha\beta}) + ({\bm q}^{-1})_{\alpha\beta} = 0, \qquad
\alpha\not=\beta.
\end{equation}

The stability of the saddle point calculation requires that the quadratic
form
\begin{equation}
\label{eq:stabil}
\delta^2(-\beta f)=
-\frac{1}{n}
\sum_{\alpha\beta}\,\Lambda'(q_{\alpha\beta})\,(\delta q_{\alpha\beta})^2 + 
        \frac{1}{n}  \mbox{Tr}({\bm q}^{-1}\, \delta {\bm q})^2
\end{equation}
must be positive definite.\cite{CriSom92}
The elements of the symmetric matrix 
 $\bm{\delta q}$ are the fluctuations $\delta q_{ab}$ 
 from the 
saddle point value (\ref{eq:spab}).

At this stage we impose the Sommers Ansatz in the DGO formalism,
i.e., we divide the matrix ${\bm q}$ into 
$n/p_0 \times n/p_0$ blocks of dimension $p_0\times p_0$ and we set
\begin{equation}
\label{eq:dgo_r0}
q_{ab}= (1-q)\delta_{ab}+(q-r)\hat \epsilon_{ab} +r
\end{equation}
where the matrix ${\bm \hat\epsilon}$ is defined as
\begin{equation}
\hat  \epsilon_{ab} = \left\{\begin{array}{ll}
         1 & \mbox{if $a$ and $b$ are in a diagonal block}\\
         0 & \mbox{otherwise} \\
        \end{array}
        \right.
\end{equation}
The Sommers solution is recovered by sending the block size $p_0$ to 
infinity with the constraint $p_0(q-r)\to -\dot\Delta$.

Details of the study of the Hessian of Eq. (\ref{eq:free2}) in the
present Ansatz are reported in Appendix \ref{app:StabSom}. Here we
concentrate on the results of that analysis relevant for the stability
of the DGO$_{R=0}$ Ansatz.  We have $n/p_0$ clusters each composed by
$p_0$ replicas. Different kinds of fluctuations can, thus, occur, e.g.
between replicas in the same cluster or in different ones, or between
clusters taken as a whole.  In the limit of the number of replicas
going to zero, the eigenvalues of the replicated Hessian matrix that
might take negative values are the following.

\begin{description}
\item{1. Fluctuations in the same cluster.}
\\
These are the fluctuations between one replica and $p_0$ others, belonging
to the same cluster. The correspondent eigenvalue is
\begin{equation}
\Lambda_1^{(1)} = - \Lambda'(q) + \frac{1}{(1-q)^2}
\end{equation}
This is the so-called {\em replicon} and describes longitudinal fluctuations
in the replica space. It must be non-negative in order
to ensure thermodynamic stability.
It
is, indeed, always positive for the Sommers solution, as 
long as $-\dot\Delta=\Delta>0$. 
To see this one just uses Eq. (\ref{eq:GammaSom}) to replace
$\Lambda'(q)$. $\Lambda_1^{(1)}$
 is the static counterpart of the dynamical stability
condition $\Gamma(\omega=0)>0$ of the Sommers solution discussed at the
end of Sec. \ref{sec:dynmod}.

\item{2. Fluctuations between clusters.}
\\
The first dangerous eigenvalue for the Sommers 
solution is, instead,
\begin{eqnarray}
\Lambda_0^{(3)} &=& -\Lambda'(r) + \frac{1}{\bigl(1-q + p_0\,(q-r)\bigr)^2}
\nonumber\\
&\phantom{=}&\hskip-20pt
               \stackrel{=}{\scriptstyle p_0\to\infty} 
                     -\Lambda'(q) + \frac{1}{\bigl(1-q - \dot\Delta)^2}
\end{eqnarray}
In this case we are considering contributions coming from fluctuations
between clusters {\em as a whole}.  Using Eq. (\ref{eq:GammaSom}), 
$\Lambda_0^{(3)}$
turns out to be always negative for $-\dot\Delta>0$, signaling the
instability that we were mentioning at the end of section
\ref{sec:dynmod}.

\item{3. Mixed Fluctuations.}
\\
 Another eigenvalue indicating an instability is
\begin{eqnarray}
 \Lambda_1^{(2)} &=& \Lambda_1^{(1)}
\\
\nonumber
      &&\hspace*{.1cm }
 - (p_0-2) \frac{q(1-q)+p_0\,(q-r)^2}{(1-q)^2\bigl(1-q + p_0\,(q-r)\bigr)^2}
\\
\nonumber
&\sim&
                     - p_0\, \frac{q(1-q)}{(1-q)^2\bigl(1-q -\dot\Delta\bigr)^2},
\quad \quad \mbox{as}~p_0\gg 1
\end{eqnarray}
It becomes infinitely large and negative as $p_0\to\infty$.  A similar
problem has been observed recently in the study of the stability of
the $R$ step DGO saddle point of the truncated model.\cite{deDomUMP}
\end{description}

>From this analysis we can conclude that the Sompolinsky theory does not 
yield  a physically consistent  static limit.


\section{The Dynamical Solution}
\label{sec:CHS-R}
The problem with the Sompolinsky solution follows from the assumption
that the width of the function $\delta_\epsilon(\omega)$, whose limit
for $\epsilon \to 0$ is a Dirac delta function, is smaller than the one of
the function $\Delta_\epsilon(\omega)$, whose limit is, instead, a
Kronecker delta, see Eq.
(\ref{eq:SompAss}). To overcome this assumption Hertz\cite{Her83}
proposed a different solution that avoids the assumption
Eq. (\ref{eq:SompAss}) by using the representations
\begin{eqnarray}
\delta_\epsilon(\omega) &=& \frac{1}{\pi}\frac{\epsilon}{\omega^2+\epsilon^2}
       = \frac{1}{\pi\omega}\,{\rm Im}\,\left[
                         \frac{\epsilon}{\epsilon - i\omega}
			                \right]
\\
\Delta_{\epsilon}(\omega) &=& \frac{\epsilon}{\epsilon-i\omega}
\end{eqnarray}
Hertz, however, assumes a standard form for the FDT and, hence, his
solution is valid only at the critical point.  CHS, studying the
dynamics of the spherical $p$-spin spin glass model, propose, instead,
a solution reproducing the correct static limit in the 1RSB phase.
The solution differs from the Sommers-Sompolinsky solution and is in
the same spirit of Hertz,\cite{Her83,Her83b} though with a different
implementation of the FDT.

We will first present the CHS solution for the $2+p$-spin model in
the Fourier space for two time sectors and we will then generalize it
to an arbitrary number of $R+1$ time sectors.
The CHS solution was originally developed (for the spherical
$p$-spin model) in the time space. 
The generalization of the CHS 
theory to $R$ time-scales in the time space, however, though
  feasible\cite{CriUMP} is quite 
tedious. Therefore, we rather work  in the $\omega$ space.


\subsection{The CHS Solution}
\begin{figure}
\includegraphics[scale=1.0]{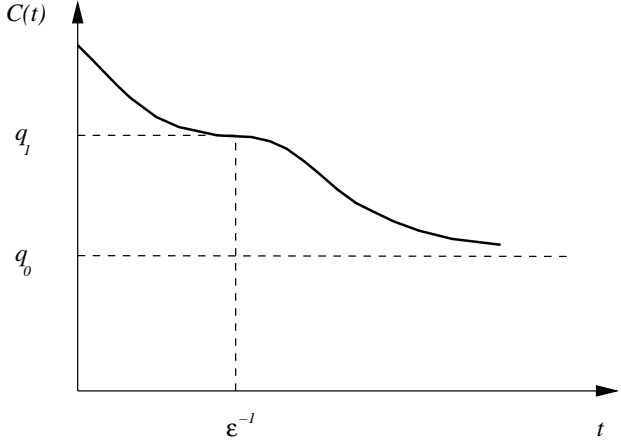}
\caption{Schematic form of correlation function with two relaxation 
         time-scales.
        }
\label{fig:corr-chs}
\end{figure}

The CHS solution assumes, in the spirit of multiple-scale analysis,
that the correlation and response functions 
are function of a {\it fast} variable $t$ and a {\it slow} variable 
$\epsilon t$, $\epsilon \ll 1$:
\begin{equation}
C(t) \Rightarrow C(t,\epsilon t), \quad 
G(t) \Rightarrow G(t,\epsilon t)
\end{equation}
The {\it fast variable} $t$ describes the decay of $C$ to the  
{\it plateau} value $q_1$ while the {\it slow variable} $\epsilon t$ 
describes the
subsequent decay to the equilibrium value $q_0$, see Fig. \ref{fig:corr-chs}.

If we are interested only in the leading order behavior for $\epsilon\to 0$,
this is equivalent to assume time-scale separation: 
either the fast variable is varying while the
slow variable is zero (i.e. the processes evolving on the long time
scale are quenched) or the slow variable is varying and 
the fast variable is infinite (i.e. the processes evolving 
on the short time-scale have already thermalized). 
Under this assumption the correlation and response
function can be represented as the sum of two separated contribution 
relative to long and short time dynamics:\cite{Note:0}
\begin{eqnarray}
C(t) &=& C_1(t) + C_0(\epsilon t) \\
G(t) &=& G_1(t) + \epsilon\,G_0(\epsilon t),
\end{eqnarray}
or
\begin{eqnarray}
C(\omega) &=& C_1(\omega) + \epsilon^{-1}C_0(\omega/\epsilon) \\
G(\omega) &=& G_1(\omega) + G_0(\omega/\epsilon),
\end{eqnarray}
where $C_1$ and $G_1$ describe the fast part and $C_0$ and $G_0$ the
slow part of $C$ and $G$, see Fig \ref{fig:corr-chs}. Alternatively,
one can employ the standard technique of multiple scale analysis,
ending up again at the leading order in $\epsilon$ with the above
expressions.\cite{CriUMP}

The functions $C_1$ and $C_0$ satisfy the boundary conditions
\begin{equation}
C_1(t=0) = 1 - q_1, \quad
C_1(t=\infty) = 0
\end{equation}
\begin{equation}
C_0(t=0) = q_1, \quad
C_0(t=\infty) = q_0
\end{equation}
where we used the spherical constraint $C(t=0) = 1$, while,
as $\epsilon\to 0$,
\begin{equation}
G_1 \not= 0\ \quad\quad\mbox{\rm iff}\ t\ll\epsilon^{-1}
\end{equation}
and
\begin{equation}
G_0 \not= 0\ \quad\quad\mbox{\rm iff}\ t\gg\epsilon^{-1}
\end{equation}
\medskip
In the regime $t\ll\epsilon^{-1}$ the FDT must be satisfied and hence 
fast parts $C_1$ and $G_1$ are related by the FDT relation 
\begin{equation}
C_1(\omega) = \frac{2}{\omega}\,{\rm Im}\, G_1(\omega)
\end{equation}
In the long-time regime $t\gg\epsilon^{-1}$ the response to an external
perturbation is given only by the degrees of freedom which 
{\it have not relaxed}, i.e. equilibrated, in the short-time regime 
$t\ll\epsilon^{-1}$, and hence only these degrees of freedom contribute
to $G_0$. 
On the other hand {\it all} degrees of freedom contribute to correlation $C_0$.
As a consequence $G_0$ cannot be related to the full $C_0$. 
If we introduce a parameter $m$, $0\leq m\leq 1$, measuring
the {\it fraction} of degrees of freedom which have not relaxed
in the short-time regime,  we have
\begin{equation}
G_0(\omega) = m\,\tilde{G_0}(\omega)
\end{equation}
where $\tilde{G}_0$ is the response function which would be observed
in the long time-regime iff {\it all} degrees of freedom would be
still active, i.e. non equilibrated.  Since all degrees of freedom
contribute to $\tilde{G}_0$ this is related to full correlation
function $C_0$ by the FDT
\begin{equation}
\label{eq:FDT00}
\tilde{C}_0(\omega) = \frac{2}{\omega}\,{\rm Im}\, \tilde{G}_0(\omega)
\end{equation}
where $\tilde{C}_0$ is Fourier transform of the non-persistent part of $C_0$:
\begin{equation}
\tilde{C}_0(t) = C_0(t) - C_0(t=\infty)= C_0(t) - q_0
\end{equation}

The equations for $m$, $q_1$ and $q_0$ are obtained by studying the 
dynamical equation in the static limit $\omega\to 0$ and $\epsilon\to 0$ 
in the two regimes $\omega\gg\epsilon$ and $\omega\ll\epsilon$.

The parameter $m$ is related to the 
discontinuity of $G(\omega)$ in passing 
from frequencies $\omega\gg\epsilon$ to frequencies $\omega\ll\epsilon$:
\begin{eqnarray}
G(\omega=0) - G(\omega_1) &=& m\,\tilde{G}_0(\omega=0) \nonumber\\
                          &=& m\,(q_1-q_0)
\end{eqnarray}
where $\omega_1$ is an infinitesimal frequency, $\omega_1\ll\Gamma_0$, 
but  goes to zero slower than $\epsilon$:
$\omega_1\gg\epsilon\to 0$.
In the
last line we used the identity $\tilde{G}_0(\omega=0) = q_1 -q_0$ which
follows from FDT relation (\ref{eq:FDT00}). Inserting now  the Dyson 
equation (\ref{eq:Dyson}) into the l.h.s of 
the above equation we end up with
\begin{equation}
\label{eq:mCHS}
m\,(q_1-q_0) = G(\omega=0)\,G(\omega_1)\,
         \bigl[\Sigma(\omega=0) - \Sigma(\omega_1)\bigr]
\end{equation}
From the expression (\ref{eq:Self}) for the self-energy it follows
\begin{eqnarray}
\Sigma(\omega=0) &-& \Sigma(\omega_1) =
\nonumber\\
&\phantom{=}&
\hskip-35pt
 \int_{0}^{\infty} dt\, \left(1-e^{-i\omega_1 t}\right)\Lambda'[C(t)]\,G_1(t)
\nonumber\\
&\phantom{=}&
\hskip-50pt
+ \epsilon \int_{0}^{\infty}dt\,\left(1-e^{-i\omega_1 t}\right)\Lambda'[C(t)]\,
G_0(\epsilon t)
\end{eqnarray}
The first integral vanishes for $\omega_1\ll\Gamma_0$, but 
$\omega_1\gg\epsilon$ as $\epsilon\to 0$. In the second integral 
only the region $t\gg\epsilon^{-1}$, where $G_0$ is different from zero, 
contributes. Therefore, we can replace $C$ with $C_0$ in the argument
of $\Lambda'[C(t)]$. Finally, by using the FDT relation (\ref{eq:FDT00}),
the leading contribution to $\Sigma(\omega=0)-\Sigma(\omega_1)$ 
for $\epsilon\to 0$ reads
\begin{equation}
\epsilon \int_{0}^{\infty}dt\, \Lambda'[C_0(\epsilon t)]\,G_0(\epsilon t) =
 m\,\bigl[\Lambda(q_1)-\Lambda(q_0)\bigr].
\end{equation}
Inserting this result into Eq. (\ref{eq:mCHS}) and using the identity
$G(\omega=0) = G_1(\omega=0) + G_0(\omega=0) = 1 - q_1 + m\,(q_1-q_0)$ 
we end up with the equation
\begin{equation}
\label{eq:q1RSB1}
\Lambda(q_1) - \Lambda(q_0) = \frac{q_1-q_0}{(1-q_1)(1-q_1 + m(q_1-q_0))}
\end{equation}

The parameter $q_0$ is the time persistent part of $C(t)$ for 
$t\gg\epsilon^{-1}$. To study this limit we consider the infinitesimal
frequency
$\omega_0 \ll\epsilon\ll\Gamma_0$ and extract the part of $C(\omega_0)$
\begin{eqnarray}
\label{eq:com0}
C(\omega_0) &=& \langle\sigma(-\omega_0)\,\sigma(\omega_0)\rangle 
\nonumber\\
            &=& G(-\omega_0)G(\omega_0)
	    \left[2\Gamma_0^{-1} + \Lambda(\omega_0)\right]
\end{eqnarray}
proportional to $\delta(\omega_0)$ for $\epsilon\to 0$.
From the form of the vertex function $\Lambda(t)$ we have, see also Eq. 
(\ref{eq:vert}),
\begin{eqnarray}
\Lambda(\omega_0) &=& 
       \int_{-\infty}^{+\infty}\, dt\, 
                     {\rm e}^{i\omega_0 t}\,
                   \bigl[\Lambda[\tilde{C}_0(\epsilon t) + q_0] - 
                           \Lambda(q_0)\bigr]
\nonumber \\
      &\phantom{=}&+ \int_{-\infty}^{+\infty}\, dt\, 
                     {\rm e}^{i\omega_0 t}\,
                           \Lambda(q_0)
\end{eqnarray}
Only the second integral contributes to the $\delta(\omega_0)$ part of 
$C(\omega_0)$. We then obtain the equation
\begin{equation}
\label{eq:q0CHS}
q_0 = G(\omega=0)^2\, \Lambda(q_0)
\end{equation}
or, equivalently,
\begin{equation}
\label{eq:q0RSB1}
\Lambda(q_0) = \frac{q_0}{\left(1-q_1+m(q_1-q_0)\right)^2}
\end{equation}

Equations (\ref{eq:q0RSB1}) and (\ref{eq:q1RSB1}) coincide the
equation for $q_1$ and $q_0$ as function of $m$ for the spherical
$2+p$ spin glass model obtained from the static replica calculation
with the Parisi 1RSB scheme, see e.g. Eq. (28) and (29) of
Ref. [\onlinecite{CriLeu06}].

\subsection{Stability of the CHS solution}
\label{ss:CHS-Stab}

The equation for the parameter $m$ follows from the dynamical
stability condition of the static limit which requires that the
$\omega\to 0$ limit of the kinetic coefficient $\Gamma(\omega)$ has to
be non negative (simply a physical requirement).  To distinguish the
two regimes $\omega\gg\epsilon$ and $\omega\ll\epsilon$ we define
$\Gamma^{-1}(\omega=0)$ as
\begin{equation}
\label{eq:kinkoe}
\Gamma^{-1}(\omega=0) = \lim_{\omega\to 0}\, \frac{i}{2\omega}
                  \bigl[G^{-1}(\omega) - G^{-1}(-\omega)\bigr]
\end{equation}
and use the frequencies $\omega_1$ and $\omega_0$ defined previously 
to perform the limit in the two regimes.

Let us first consider the frequency $\omega\gg\epsilon$.
In this case the limit $\omega\to 0$ must be 
evaluated as 
\begin{equation}
\lim_{\omega\to 0} f(\omega) := 
\lim_{\omega\to 0} \lim_{\epsilon\to 0} f(\omega) = f(\omega_1)
\end{equation}
Since $G_0(\omega_1/\epsilon)\simeq 0$, we, thus, have
\begin{eqnarray}
G^{-1}(\omega_1)-G^{-1}(-\omega_1) &=&
-2i\omega_1\bigl[\Gamma_0^{-1}  + A_1(\omega_1)\bigr] 
\\
&\phantom{=}&
    -\Lambda'(q_1)\bigl[G_1(\omega_1) - G_1(-\omega_1)\bigr]
\nonumber
\end{eqnarray}
where $A_1(\omega_1)$ is a finite and positive quantity.
Inserting this expression into Eq. (\ref{eq:kinkoe})
we end up with\cite{Note:Gamma}
\begin{equation}
\Gamma^{-1}(\omega_1) = \frac{\Gamma_0^{-1} + A_1(\omega_1)}
                               {1 - G_1(\omega=0)^2\Lambda'(q_1)}
\end{equation}
so that the requirement $\Gamma(\omega_1)\geq 0$ leads to the condition
\begin{equation}
\label{eq:stb1CHS}
1 - G_1(\omega=0)^2\Lambda'(q_1) = 
1 - (1-q_1)^2\Lambda(q_1) \geq 0
\end{equation}

A similar calculation for $\omega\ll\epsilon$, i.e., evaluating now the
limit $\omega\to 0$ as
\begin{equation} 
\lim_{\omega\to 0} f(\omega) := 
\lim_{\epsilon\to 0} \lim_{\omega\to 0} f(\omega) = f(\omega_0)
\end{equation}
leads to 
\begin{equation}
\label{eq:stab0}
\Gamma^{-1}(\omega_0) = \frac{\Gamma_0^{-1} + A_2(\omega_0)}
                               {1 - G(\omega=0)^2\Lambda'(q_0)}
\end{equation}
where $A_2(\omega_0)$ is finite and positive. 
We have, thus, the second 
condition for the stability:
\begin{eqnarray}
\label{eq:stb0CHS}
1 &-& G(\omega=0)^2\Lambda'(q_0) = \nonumber\\
&\phantom{-}& 1 - (1-q_1+m(q_1-q_0))^2\Lambda(q_0) \geq 0
\end{eqnarray}

The dynamical stability conditions of the static limit,
 Eqs. (\ref{eq:stb1CHS}) and (\ref{eq:stb0CHS}), coincide with the
 stability conditions of the 1RSB saddle point computed in the static
 replica calculation of Ref.  [\onlinecite{CriLeu06}] [Eqs. (31) and
 (32)].

>From a dynamical point of view, and for the consistency of the
calculation, we must require that $\Gamma(\omega_1) = 0$. Indeed, if
$\Gamma(\omega_1)>0$ the correlations decay exponentially fast and the
system equilibrates on a time-scale of order
$\Gamma^{-1}(\omega_1)$. This would imply that all degrees of freedom
have relaxed before entering the regime $t\gg\epsilon^{-1}$ and $m=0$,
i.e., we have back the RS solution.  To yield
a two time scales solution, then, the condition (\ref{eq:stb1CHS})
 becomes the additional equation:
\begin{equation}
\label{eq:marg}
1 - (1-q_1)^2\Lambda(q_1) = 0
\end{equation}
the so called {\it marginal condition}.\cite{Note:Marginal} The
self-consistency equations (\ref{eq:q1RSB1}), (\ref{eq:q0RSB1}),
(\ref{eq:marg}) and the stability condition (\ref{eq:stab0}) yield the
CHS dynamical solution of the spherical $2+p$ model in the 1RSB phase.
The dynamic RS/1RSB transition line in Fig. \ref{fig:static_pd} can be
obtained right by Eq. (\ref{eq:marg}) as the curve where $q_1$ jumps
discontinuously from zero to a finite value.

Summing up, the CHS solution presents -- in the 1RSB phase -- an
infinite time limit different from the static solution.  Indeed, it
coincides with the solution known as ``dynamic'', where the 1RSB phase
nucleates at higher free energy than the equilibrium one.  The stable
phase, in the sense of lower free energy, in this regime is still the
RS one but a 1RSB phase exists, is locally stable, and despite a
higher free energy it dominates the dynamics due to the very large
degeneracy of the metastable states belonging to it.  In other words,
in its evolution on the free energy surface, the system will find
itself with probability one in a local minimum of the 1RSB solution
simply because the number of these minima is exponentially large, in
the system size, with respect to the number of global
minima of the RS solution.  The
logarithm of the number of equivalent minima is what is called the
{\it complexity}, and hence, the dynamics of the system is dominated by the
solution with the largest complexity.  The marginal condition
Eq. (\ref{eq:marg}) is, indeed, nothing else than the condition of
maximal complexity in the static Parisi replica theory.\cite{CriLeu06}
The static solution corresponds, instead, to the lowest minima of the
free energy and has vanishing complexity. This is the reason
why the two solutions differ.

To be more explicit, for the $2+p$-spin model, in
Fig. \ref{fig:static_pd} we noticed two lines between the RS and the
1RSB phases. The dotted one is the line at which a 1RSB solution (with
a $q_1>0$) arises, even though the statics stays RS. This is the
dynamic phase transition that, as we have just seen above,
 is also obtained from the static limit
of the CHS solution.  The solid line marks the thermodynamic
transition to a stable 1RSB phase. In the 1RSB region an extensive
complexity exists, monotonically increasing between the lowest
equilibrium free energy (at which it is zero) and the threshold free
energy (where it takes its maximum value). The infinite time limit of
the CHS solution describes those states lying at the threshold free
energy. 

As Eq. (\ref{eq:marg}) cannot be satisfied anymore with $0 < q_1 < 1$
and $\Gamma(\omega_1)$ becomes negative, 
 a different solution is needed, involving more time-scales and, 
correspondingly, more overlap order parameters. This is right 
the generalization that  we are going to analyze in the next section.

In particular, in the $T-J_p$ diagram in Fig.  \ref{fig:static_pd},
the limit of validity of Eq. (\ref{eq:marg}) is represented by the
dynamic transition line between the 1RSB and the 1-FRSB phase. In that case
the number of scales required to stabilize the solution out of the
region of validity of the 1RSB phase (for increasing $T$ at fixed
$J_p$ or decreasing $J_p$ at fixed temperature) becomes a continuous
set, plus an apart step relative to the shortest time-scales.

We stress that $\Gamma(\omega_0)$ remains non-negative since it
describes the relaxation of the systems to equilibrium
for $t\gg\epsilon^{-1}$. 
In Appendix \ref{app:Dyn2p} we show in detail why for the $2+p$ spherical model
the marginal condition on the intermediate time scale is necessary in order to
guarantee relaxation to equilibrium on the longest time scale.
In the $2+p$ spherical model the instability appears on the intermediate 
time scale, however, in general, it may appear on the longest time
scale as well, with a negative $\Gamma(\omega_0)$. In the present scenario 
this means that a new (infinite) time-scale enters into the game and must 
be included into the description of the dynamics.

\subsection{The $R$-time-scale CHS Solution}
\begin{figure}
\includegraphics[scale=1.0]{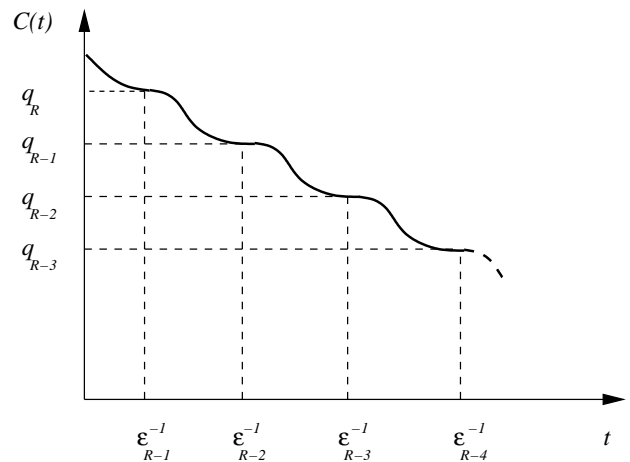}
\caption{Schematic form of correlation function with many relaxation 
         time-scales in CHS theory.
        }
\label{fig:corr-e}
\end{figure}

The extension of the CHS theory to the case of $R$ different diverging 
relaxation time scales follows the same initial steps of the Sompolinsky 
solutions, namely one introduces $R$ long time-scales, 
see Fig. \ref{fig:corr-e},
\begin{equation}
\Gamma_0^{-1}\ll\epsilon_{R-1}^{-1}\ll\epsilon_{R-2}^{-1}\ll\cdots
\ll\epsilon_1^{-1}\ll\epsilon_0^{-1}
\end{equation}
all of which eventually diverge in the prescribed order in the thermodynamic 
limit.\cite{Note1}
In what follows we shall denote with ${\bm \epsilon}$ the set of the 
$R$ frequencies $\epsilon_r$ and assume that 
the limit ${\bm \epsilon}\to \bm 0$ is always taken in order, 
i.e., 
\begin{equation}
\lim_{{\bm \epsilon}\to 0} := \lim_{\epsilon_{R-1}\to 0}\cdots
                               \lim_{\epsilon_0\to 0}
\end{equation}
A convenient way of studying the dynamics in the limit ${\bm
\epsilon}\to \bm 0$ is by using the multiple scale analysis.  One then
assumes that the correlation function $C(t)$, as well as the response
$G(t)$, is a function of the fast variable $\tau_R = \epsilon_R t$ with
$\epsilon_R = \Gamma_0$ and $R$ slow variables $\tau_r =
\epsilon_r\,t$ ($r=0,\ldots,R-1$) describing the motion in each time
sector $\epsilon_r^{-1}\ll t\ll\epsilon_{r-1}^{-1}$:
\begin{equation}
  C(t) \Rightarrow C(\tau_R,\tau_{R-1},\ldots,\tau_1,\tau_0).
\end{equation}

The leading behavior for ${\bm \epsilon}\to \bm 0$ in the time sector
$\epsilon_r^{-1}\ll t\ll\epsilon_{r-1}^{-1}$ is obtained by assuming
that only processes evolving on times $t\gg\epsilon_{r}^{-1}$ but
$t\ll\epsilon_{r-1}^{-1}$ contribute (i.e. $\tau_r=$ finite) whereas
those evolving on slower time scales are quenched ($\tau_{s<r} = 0$)
and those evolving on faster time scale are thermalized ($\tau_{s>r} =
\infty$).  Under this assumption of time scale separation, $C(t)$ can
be represented as the sum of $R+1$ distinct terms
\begin{equation}
C(t) = \sum_{r=0}^{R}\,C_r(\tau_r),\quad \quad\tau_r = \epsilon_r\,t, 
\end{equation}
one for each sector. 

The functions $C_r$ satisfy the normalization condition
(spherical constraint)
\begin{equation}
C(t=0) = \sum_{r=0}^{R}\,C_r(\tau_r=0) = 1.
\end{equation}
We can now
 split off the $r$-sector function $C_r$ and  take  the
limit ${\bm \epsilon}\to \bm 0 $ with $\tau_r $ finite. 
Taking the limit $\tau_r\to\infty$, afterward, so that  $C(t)\to q_r$,
cfr. Eq. (\ref{eq:qr}), we have the additional conditions:
\begin{equation}
\sum_{s=0}^{r-1}\,C_s(\tau_s=0) +
       \sum_{s=r}^{R}\,C_s(\tau_s=\infty) = q_r \quad \forall r=0,\ldots,R
\end{equation}
It is useful to define for each sector the non-persistent part of the 
correlation function as
\begin{equation}
\tilde{C}_r(\tau) = C_r(\tau) - C_r(\tau=\infty)
\end{equation}
so that the above conditions become
\begin{equation}
\tilde{C}_r(\tau=0) = q_{r+1} - q_r, \quad
\tilde{C}_r(\tau=\infty) = 0
\quad \forall r=0,\ldots,R
\end{equation}
with $q_{R+1} = 1$, while 
the whole $C(t)$ reads
\begin{equation}
\label{eq:crtCHS}
C(t) = \sum_{r=0}^{R}\tilde{C}_r(\tau_r) + q_0.
\end{equation}

By similar arguments we obtain the following representation for the response 
function $G(t)$
\begin{equation}
\label{eq:grtCHS}
G(t) = \sum_{r=0}^{R}\, \epsilon_r\,G_r(\tau_r)
\end{equation}
where each function $G_r$ varies only in the corresponding sector $r$,
where $\tau_r\sim O(1)$ for ${\bm \epsilon}\to \bm 0$, and vanishes in
all sectors with $s<r$.  The function $G_r$ represents the response of the
system to a perturbation in the time sector labeled by $r$, i.e., the
response due to all degrees of freedom which have not equilibrated in
previous sectors. As a consequence, $G_r$ cannot be related to the {\it
full} correlation function $\tilde{C}_r$ since all degrees of freedom,
equilibrated or not, contribute to the latter.  By introducing the
parameter $0\leq m_{r+1}\leq 1$ as the fraction of degrees of freedom
which have {\it not relaxed up to sector} $r+1$, and hence do
contribute to the response in the next sector $r$, we can write
\begin{equation}
G_r(\tau) = m_{r+1}\,\tilde{G}_r(\tau)
\end{equation}
where $\tilde{G}_r$ is the {\it full} response in sector $r$ due
to {\it all} degrees of freedom, equilibrated and not:
\begin{equation}
\label{eq:FDTr}
\tilde{C}_r(\omega) = \frac{2}{\omega}\,{\rm Im}\, \tilde{G}_r(\omega)
\end{equation}
By definition $m_{R+1} = 1$, since in the first sector all degrees of freedom 
contribute to the response, while $m_0=0$ since the system equilibrates in 
the last sector.

By taking the Fourier transform of Eqs. (\ref{eq:crtCHS}) and 
(\ref{eq:grtCHS}) we have
\begin{eqnarray}
\label{eq:comCHS}
C(\omega) &=& \sum_{r=0}^{R}\epsilon_{r}^{-1}\tilde{C}_r(\omega_r/\epsilon_r) 
              +  2\pi\,q_0\delta(\omega) \\
\label{eq:gomCHS}
G(\omega) &=& \sum_{r=0}^{R}\,G_r(\omega_r/\epsilon_r).
\end{eqnarray}

As for the CHS$_{R=1}$ solution, the equations for $q_r$ and $m_r$ are
obtained by studying the static limit $\omega\to 0$ separately in each
sector.  We, then, introduce the set of infinitesimal frequencies
$\omega_r$, with $\omega_r\ll\epsilon_r$ but
$\omega_r\gg\epsilon_{r-1}$, all of which go to zero as
${\bm\epsilon}\to 0$, so that the $\omega\to 0$ limit in sector $r$
just reads
\begin{eqnarray}
\lim_{\omega\to 0} f(\omega) :&=& \lim_{\epsilon_{R-1}\to 0}\cdots
  \lim_{\epsilon_r\to 0}\lim_{\omega\to 0}\lim_{\epsilon_{r-1}}\cdots
  \lim_{\epsilon_0\to 0}\, f(\omega) \nonumber\\
  &=& f(\omega_r)
\end{eqnarray}

The parameter $q_0$ is the singular part of $C(\omega_0)$,
see Eq. (\ref{eq:comCHS}), and repeating the steps from 
Eq. (\ref{eq:com0}) to Eq. (\ref{eq:q0CHS}) we end up with 
\begin{equation}
q_0 = G(\omega=0)^2\,\Lambda(q_0).
\end{equation}
where $G(\omega=0)$ must be evaluated from the expression (\ref{eq:gomCHS}):
\begin{eqnarray}
G(\omega=0) &=& \sum_{r=0}^{R}\, m_{r+1}\,\tilde{G}_r(\omega=0) \nonumber\\
    &=& 1 - q_R + \sum_{r=0}^{R-1} m_{r+1}(q_{r+1}-q_r) \nonumber\\
    &=& 1 - q_R + \sum_{r=0}^{R} m_r(q_r-q_{r-1}) \nonumber \\
    &=& \chi_0
\end{eqnarray}
[cf. Eq. (\ref{eq:chiP})]. Here we have used the relations
$\tilde{G}_r(\omega=0) = q_{r+1} - q_r$, following from the FDT relation 
(\ref{eq:FDTr}), and $m_0=0$. Thus we have
\begin{equation}
\label{eq:q0CHSr}
\Lambda(q_0) = \frac{q_0}{\chi_0^2}
\end{equation}
coinciding with Eq. (\ref{eq:q0Par}) obtained from the static
replica calculation within the Parisi RSB scheme. 
The presence of an external field $h$ would,  indeed, just add a term 
$-(\beta h)^2$ to the r.h.s. of this equation, as can be easily verified.

To find the equation for $q_r$ with $r=1,\ldots, R$, we consider the
discontinuity of $G(\omega)$ in passing from one sector to the next one:
$G(\omega_{r-1})-G(\omega_r)$. By observing that
$G_s(\omega_r/\epsilon_s)\simeq 0$ for $s<r$ (since
$\omega_r/\epsilon_s\gg 1$)
while 
$G_s(\omega_r/\epsilon_s)\to G_s(\omega=0)$ for $s\geq r$ (since
$\omega_r/\epsilon_s\ll 1$), it follows that
\begin{eqnarray}
G(\omega_r) &=& \sum_{s=0}^{r-1}G_s(\omega_r/\epsilon_s) 
             + \sum_{s=r}^{R} G_s(\omega_r/\epsilon_s) \nonumber\\
            &=&  \sum_{s=r}^{R}G_s(\omega=0) \nonumber\\
	      &=& \sum_{s=r}^{R}m_{s+1}(q_{s+1}-q_s)
\end{eqnarray}
The difference $G(\omega_{r-1}) - G(\omega_{r})$, therefore, reads
\begin{equation}
 G(\omega_{r-1}) - G(\omega_r) = m_r\,(q_r - q_{r-1}).
\end{equation}
that, using the Dyson equation (\ref{eq:Dyson}), can be transformed into
\begin{equation}
\label{eq:qrCHSr}
m_r\,(q_r-q_{r-1}) = G(\omega_r)\,G(\omega_{r-1})
             \bigl[\Sigma(\omega_{r-1}) - \Sigma(\omega_r)\bigr].
\end{equation}
This relation is valid for  $r=1,\ldots, R$. For $R=1$ it reduces 
Eq. (\ref{eq:mCHS}) of the CHS solution.

For the spherical $2+p$ model the difference
$\Sigma(\omega_{r-1}) - \Sigma(\omega_r)$ can be easily evaluated:
from Eq. (\ref{eq:Self}) and the definition of $\epsilon_r$ we have
\begin{eqnarray} 
\Sigma(\omega_{r-1}) - \Sigma(\omega_r) &\phantom{=}& \nonumber\\
&\phantom{=}&
\hskip-80pt =
 \sum_{s=0}^R\epsilon_s\int_{0}^{\infty}dt\,
 \left(e^{i\omega_{r-1}t} - e^{i\omega_{r}t}\right)\Lambda'[C(t)]\,
  G_s(\epsilon_s t) 
\nonumber\\
&\phantom{=}&
\hskip-80pt = \epsilon_{r-1}
\int_{0}^{\infty}dt\,
 \left(e^{i\omega_{r-1}t} - e^{i\omega_{r}t}\right)
\Lambda'[C(t)]\,  G_{r-1}(\epsilon_{r-1}t)
\nonumber\\
&\phantom{=}&
\hskip-80pt =
\int_{0}^{\infty}d\tau\,
\Lambda'[\tilde{C}_{r-1}(\tau) + q_{r-1}]\,G_{r-1}(\tau)
\nonumber\\
&\phantom{=}&
\hskip-80pt =
m_r\,\bigl[\Lambda(q_r)-\Lambda(q_{r-1})\bigr]
\end{eqnarray}
where we have used the fact that
 only the term with $s=r-1$ yields a finite contribution for 
${\bm\epsilon}\to 0$.

With this expression for the difference 
$\Sigma(\omega_{r-1}) - \Sigma(\omega_r)$, and using the identity 
[cfr. Eq. (\ref{eq:chiP})]
\begin{eqnarray}
\label{eq:G_r-1}
G(\omega_{r-1}) &=& 
   \sum_{s=r-1}^{R}\, m_{s+1}\,\tilde{G}_s(\omega=0) \nonumber\\
    &=& 1 - q_R + \sum_{s=r-1}^{R-1} m_{s+1}(q_{s+1}-q_s) \nonumber\\
    &=& 1 - q_R + \sum_{s=r}^{R} m_s(q_s-q_{s-1}) \nonumber \\
    &=& \chi_r
\end{eqnarray}
we finally obtain the equation for the static limit of the $R$
time scale CHS solution:
\begin{equation}
\label{eq:qrPara}
\Lambda(q_r) - \Lambda(q_{r-1}) = \frac{q_r - q_{r-1}}
       {\chi_r\, \chi_{r+1}}, \quad
r = 1,\ldots, R
\end{equation}
coinciding with the result from the static replica calculation, see Eq. 
(\ref{eq:qrPar}).

\subsection{Stability of the $R$-time CHS solution}

As for the CHS$_{R=1}$ solution the equation for $m_r$ follows from
the stability condition of the static limit $\omega\to 0$ in sector r.
From the definition (\ref{eq:kinkoe}) of the kinetic coefficient
$\Gamma(\omega)$ we have
\begin{eqnarray}
\Gamma^{-1}(\omega_r) &=& \frac{i}{2\omega_r}\left[
  G^{-1}({\omega_r}) - G^{-1}({-\omega_r}) \right]
\nonumber\\
&=& \Gamma_0^{-1} - \frac{i}{2\omega_r}\left[
    \Sigma^{-1}({\omega_r}) -  \Sigma^{-1}({-\omega_r})\right]
\nonumber\\
\end{eqnarray}
For the spherical $2+p$ model the difference 
$\Sigma(\omega_r)-\Sigma(-\omega_r)$ is given by
\begin{eqnarray} 
\Sigma(\omega_r) - \Sigma(-\omega_r) &\phantom{=}& \nonumber\\
&\phantom{=}&
\hskip-75pt = 
\int_{0}^{\infty}dt\,
 \left(e^{i\omega_rt} - e^{-i\omega_{r}t}\right)
\Lambda'[C(t)]\,  G(t)
\nonumber\\
&\phantom{=}&
\hskip-75pt =
\tilde{\Sigma}(\omega_r) - \tilde{\Sigma}(-\omega_r) + 
\Lambda'(q_r)\bigl[G(\omega_r) - G(-\omega_r)\bigr]
\nonumber\\
\end{eqnarray}
where
\begin{equation}
\tilde{\Sigma}(\omega_r) =
 \int_{0}^{\infty} dt\, e^{i\omega_r t}
\bigl[\Lambda'[C(t)] - \Lambda'(q_r)\bigr]\,G(t).
\end{equation}
As a consequence we have
\begin{equation}
\Gamma^{-1}(\omega_r) = \frac
       {\Gamma_0 - i\frac{\partial}{\partial\omega_r}\tilde{\Sigma}(\omega_r)}
       {1-G(\omega_r)^2\,\Lambda'(q_r)}.
\end{equation}
The quantity $i(\partial/\partial\omega_r)\tilde{\Sigma}(\omega_r)$ is
real and negative, therefore, the requirement
$\Gamma^{-1}(\omega_r)\geq 0$ leads to the dynamical stability
condition
\begin{equation}
1 - G(\omega_r)^2\,\Lambda'(q_r)\geq 0
\end{equation}
which can be written [cfr. Eq. (\ref{eq:G_r-1})] as
\begin{equation}
\label{eq:mCHSr}
-\Lambda'(q_r) + \frac{1}{\chi_{r+1}^2}\geq 0
\end{equation}
where $\chi_{r}$ is defined in Eq. (\ref{eq:chiP}).
We recover then the condition for stability of the $R$-RSB saddle
point in the replica calculation.\cite{CriLeu06}

At difference with the static calculation, however, the dynamical
solution requires that all $\Gamma^{-1}(\omega_r)$, but the last one
for $r=0$, vanish.
Indeed, as discussed for the CHS solution, if  it happens that
$\Gamma^{-1}(\omega_r)>0$ for some $r=1,\ldots, R$ then all degrees of
freedom not yet thermalized up to sector $r$ relax in the sector $r$,
 so that $m_s=0$ for $s\leq r$. This, in turn,
implies that the number of diverging relaxation time scale is
$r<R$, and not $R$ as initially
assumed.\cite{Note:Marginal} This argument does
not apply to the last sector $r=0$. Indeed, by assumption, the system
equilibrates in this sector and this
is feasible only if $\Gamma^{-1}(\omega_0)$ is positive.\cite{Note-stab}

If the above requirements cannot be satisfied and one or more 
$\Gamma(\omega_r)$ are negative, including the last sector $r=0$,
then additional time scale(s) have to be included into the description.
This is what happens, e.g., in the $2+p$ spherical model in the
1-FRSB and FRSB phases. Actually in these phases an infinite number of 
successive time scales is required in order to stabilize the dynamics. 
Nevertheless it can be seen that increasing $R$ the violation of 
dynamical (marginal) stability decreases and vanishes as $R\to\infty$.

We stress that similarly to what happens for the CHS solution in the 1RSB 
phase, also the extension of the CHS theory to $R$-time-scales does not 
reproduce exactly the {\it same}
static solution found from the static replica calculation.
Indeed, while the equations (\ref{eq:q0CHSr}) and (\ref{eq:qrPara}) 
for $q_0$ and $q_r$ and the condition $\Gamma(\omega_0)>0$ are the same as 
those found from the static replica calculation, the equations
\begin{equation}
\label{eq:gamma-marg}
\Gamma(\omega_r) = 0, \quad \mbox{ for}\ r = 1,\ldots, R
\end{equation}
differ from the corresponding ones in statics.  For any finite $R$ we
have, thus, the same phenomenon already observed for systems described
by $1$-RSB solution.\cite{KirWol87,CriSom95} As in that case, the
difference between the static free energy and the free energy of the
static limit of the dynamical solution corresponds to the existence of
an extensive complexity of the $R$-RSB solution, i.e., to the presence of
a macroscopic number of statistically equivalent metastable states
dominating the dynamics.  In the free energy landscape describing the
phase space of the system, such states are at a free energy level
larger then the free energy of the static minimum, nevertheless they
dominate the dynamics due to their macroscopic number.  In the
mean-field picture we are adopting here, for $\epsilon_r\to 0$, the
system is stuck in these threshold states because of the consequent
decoupling between processes at different time scales.  Relaxing such
constraint, i.e., going beyond mean-field, the evolution from the
threshold states at a given step of the RSB can be, instead,
allowed.\cite{CriRit00}

The difference disappears in the FRSB phase where the complexity 
vanishes. Indeed defining 
$m_r = x(q_r)$ from either Eq. (\ref{eq:qrPara}) or (\ref{eq:gamma-marg}) 
we obtain in the limit $R\to\infty$ 
\begin{equation}
\frac{d}{dq}\Lambda(q) = \frac{1}
        {\left(1-q_R +\int_{q}^{q_R}dq'\ x(q')\right)^2}
\end{equation}
which is the FRSB solution of the spherical $2+p$ model, cfr. Eq. (53) of
Ref. [\onlinecite{CriLeu06}].
It is easy to show that in the $R\to\infty$ limit the order parameter 
function $q(x)$ satisfies the Parisi anti-parabolic differential equation.
We also note that at difference with the Sompolinsky theory the 
$R$-time-scale CHS solution does not introduce the additional function 
$\Delta(x)$.

\section{Summary and Conclusions}

We have addressed the problem of the equilibrium dynamics of spin
glass systems. One of the issues that makes equilibrium dynamics worth
to be studied is its connection with the static properties of the
systems, i.e., those obtained from statistical mechanics via the
partition function.  While the statistical mechanics of spin glass
systems is well developed, the equilibrium dynamics is less known. The
usual theory for equilibrium dynamics of spin glass systems is the
Sompolinsky theory that in the FRSB phase leads to a static solution
in agreement with the statistical mechanic one, provided one imposes
the Parisi gauge $d\Delta(x)/dx = -x\,dq(x)/dx$.  The Sompolinsky
theory has received further support from de Dominicis, Gabay and
Orland (DGO) who, using a replica symmetry breaking scheme with two
order parameters (a Parisi-like overlap $q$ and a Sompolinsky-like
anomaly $\Delta$), derived the FRSB Sompolinsky solution from
equilibrium statistical mechanics.  Despite these results the
Sompolinsky theory was object of criticisms and its validity is still
not well established.

In this work we have analyzed in details the Sompolinsky solution
using the spherical $2+p$ spin glass model.  The main motivations in
using this model are (i) that its static solution is completely
exactly known and (ii) that, besides displaying a FRSB phase, it possesses
stable 1RSB and 1-FRSB phases, so that - unlike in the SK model - we
can test the Sompolinsky solution in phases other than the FRSB.

The first result of our study is that if the number $R$
of relaxation times (equivalently the number of replica 
symmetry breaking steps in the static equivalent DGO description)
is finite then the Sompolinsky theory leads to a static solution that
cannot be traced back to the static solution obtained with the Parisi 
RSB scheme. The two solutions can coincide only in the 
FRSB phase, when $R\to\infty$ and $q(x)$ becomes a continuous function,
provided one fixes the Parisi gauge.

To understand the properties of the Sompolinsky solution we have
studied it in the 1RSB phase of the $2+p$ model.  The analysis,
performed within the equivalent DGO theory, reveals that the
fluctuations about the DGO saddle point yielding the static limit of
the Sompolinsky solution not only have negative eigenvalues, but
some of them go to minus infinite. The
saddle point is, therefore, unstable and the Sompolinsky solution in
the infinite time limit is not a physically consistent solution.  As
already noted by Hertz\cite{Her83} the weak point of Sompolinsky
theory is in the assumption that each time sector $r$ is assumed to
contribute to the effective static field $H(\{z\})$ in the spin
equation of motion with the full magnetization $\bar m_r(\{z\})$
induced at that time scale by the slow noise $z$, mathematically
expressed by Eq. (\ref{eq:SompAss}).

In the second part of the paper we have presented an alternative
theory for the equilibrium dynamics of spin glass systems. The theory,
based on the CHS solution of the spherical $p$-spin spin glass model,
differs from the Sompolinsky theory in that it uses a modified form of
the FDT theorem to deal with the anomalous contribution to the
response function, overcoming the Sompolinsky assumption Eq.
(\ref{eq:SompAss}). In this theory, in the static limit, the
parameters are $q_r$, the time persistent part of the correlation
function at (infinite) time scale $t_r$, and $m_r$, the fraction of
non equilibrated degrees of freedom at scale $t_r$ entering into a
modified FDT. No anomaly functions are introduced to represent the
zero field cooled static susceptibility.

The equations for $q_r$ have the same functional form of those derived 
from statistical mechanics using the Parisi RSB Ansatz.
For any finite $R$, however, the equations for $m_r$ have a 
different form. The reason is that in the dynamic theory the equations 
for $m_r$ follow from the condition that the dumping function 
must be zero ({\em marginal condition}) for all  but the longest
time scale:\cite{Note-stab}
\begin{equation}
\Gamma(\omega_r) = 0,\quad  r = 1,\ldots, R.
\label{eq:marginal_dyn}
\end{equation}
In the static replica calculation the self-consistency equations for
$m_r$ are, instead, obtained by the stationarity of the replicated
free energy functional with respect to variation of $m_r$, i.e., from
the vanishing of the derivative of the replica free energy with
respect to $m_r$.  From the replica calculation point of view, on the
contrary, the dynamic marginal condition corresponds to the
requirement of a maximal derivative of the free replica free energy
with respect to $m_r$, that is maximal
complexity.\cite{KirWol87,CriHorSom93,CriSom95,Nie95}

The difference between dynamic and static solutions is due to the
 degeneracy of the metastable excited states that yields an extensive
 complexity at free energies higher than the static one. Even though
 their weight is smaller than the one of the equilibrium state, the
 states at higher free energy (``threshold states'') are statistically
 much more relevant (their number is exponentially larger as the size
 of the system increases) and, therefore, a system cooled down from
 high temperature will end up in one of these with probability one.
 Because of the mean field nature of the models considered and the
 consequent growth of barriers with the size, the system cannot evolve
 anymore out of the threshold states in a relaxation dynamics and the
 equilibrium states become unreachable in the thermodynamic limit.
 This might be possibly bypassed considering timescales that are not
 completely decoupled. In our notation it would amount to use non
 vanishing ${\bm \epsilon}$ values and compute the first correction to
 the leading behavior for ${\bm \epsilon}\to 0$, not an easy task.

When more RSB steps are considered, the complexity depends on more 
breaking parameters $m_r$ and the threshold value is obtained by maximizing
the complexity with respect to all of them. In the dynamical formalism it is 
 equivalent to impose Eq. (\ref{eq:marginal_dyn}). 
This selects the ensemble of statistically
equivalent minima of the (exponentially) more numerous kind, that is, those
at higher level in the free energy corrugated landscape.
As the number of steps is increased, the complexity function counting the 
number of minima decreases, as well as the difference between the 
 dynamic (threshold) free energy and the static 
(equilibrium) one.\cite{Note:MarStates}
In the
limit where the stable phase is FRSB this difference eventually reduces to zero, as
e.g. in Ref.  [\onlinecite{CriLeuParRiz04}] for the case of the Ising
SK model.  The same effect can be detected in the Ising $p$-spin
model\cite{MonRic03} at zero temperature passing from a 1RSB to a 2RSB
Ansatz, even though both solutions are physically inconsistent even
at the static point.  The advantage of the spherical $2+p$ with
respect to the above mentioned models is that three 
apart spin-glass stable phases exist, each obtained by a different -- physically
consistent -- RSB solution, where the above considerations have been tested.

\begin{acknowledgments}
The Authors wish to thank C. de Dominicis for all the discussions which 
pushed us on this subject. A.C. also thanks the SPTH of CEA in Saclay
where part of this work was done, for the very kind hospitality and support.

\end{acknowledgments}


\appendix
\section{The De Dominicis, Gabay, Orland (DGO) solution}
\label{app:DGO}
In this appendix we sketch the derivation of the De Dominicis, Gabay, 
Orland\cite{deDomGabOrl81} (DGO) static solution for the $2+p$ spin glass 
model and show its relation with the Sompolinsky solution.

In the replica approach the static solution for the spherical $2+p$ spin glass 
model is given by the $n\to 0$ limit, where $n$ is the number of replica,
of stationary point of the replica functional
$f[Q_{ab},\Lambda_{ab}]$:\cite{CriSom92,CriLeuPar02}
\begin{eqnarray}
-n\beta f\left[Q_{\alpha\beta},\Lambda_{\alpha\beta}\right] &=&
             \frac{1}{2}\sum_{\alpha\beta}^{1,n}g\left(Q_{\alpha\beta}\right)
      - \frac{1}{2}\sum_{ab}^{1,n}\Lambda_{\alpha\beta}Q_{\alpha\beta}
\nonumber \\
&\phantom{=}&
\hskip -78pt
+\log\mbox{Tr}_{\sigma}
\exp\left(\frac{1}{2}\sum_{\alpha\beta}\Lambda_{\alpha\beta}\,
   \sigma_{\alpha}\sigma_{\beta} +
            b\sum_{\alpha}\sigma_\alpha\right)
\label{eq:fen_Qab}
\end{eqnarray}
where $Q_{\alpha\beta}$ is the spin-overlap matrix in the replica space 
between replicas $\alpha$ and $\beta$,
$\Lambda_{\alpha\beta}$, the Lagrange multiplier associated with 
$Q_{\alpha\beta}$ and $b = \beta h$ the external field.
The function $g(x)$ is defined as $dg(x)/dx = \Lambda(x)$, with 
$\Lambda(x)$ given by Eq. (\ref{eq:Lambda}).
Moreover for the spherical model
\begin{equation}
\Lambda_{\alpha\alpha} = \overline{\lambda} 
\end{equation}
is the Lagrange multiplier fixing the spherical constraint
$Q_{\alpha\alpha} = 1$, and 
\begin{equation}
\mbox{Tr}_{\sigma} := \prod_{\alpha}\int_{-\infty}^{+\infty}\,d\sigma_\alpha
\end{equation}

The $R$-step DGO solution is obtained by taking the $n\times n$  matrix 
$Q_{\alpha\beta}$ made of $(n/p_0)^2$ sub-matrices  $q_{ab}$ and $r_{ab}$ of 
size $p_0\times p_0$
\begin{equation}
Q_{\alpha\beta} = \left( 
    \begin{array}{c|c|c}
      q_{ab} & r_{ab} & r_{ab} \\
   \hline
      r_{ab} & q_{ab} & r_{ab} \\
   \hline
      r_{ab} & r_{ab} & q_{ab} \\
    \end{array}
\right)
\label{app:SomMatr}
\end{equation}
with each matrix $q_{ab}$ and $r_{ab}$ an $R$-RSB Parisi matrix:
\begin{eqnarray}
q_{ab} = \sum_{t=0}^{R+1} (q_t - q_{t-1})\,\prod_{k=0}^{t-1}\delta_{a_k,b_k}
\\
r_{ab} = \sum_{t=0}^{R+1} (r_t - r_{t-1})\,\prod_{k=0}^{t-1}\delta_{a_k,b_k}
\end{eqnarray}
where 
\begin{eqnarray}
a_k &=& 0,\ldots,p_k/p_{k+1}-1
\quad \mbox{with}
\\ 
\nonumber 
&&\ \ 1=p_{R+1}<p_R<\cdots<p_1<p_0,
\\
\nonumber
q_{-1}&=&r_{-1} =0,
\\
\nonumber 
q_{R+1} &=& \overline{q} = 1  \quad \mbox{(for the spherical constraint)}
\\
\nonumber
r_{R+1}&=&r_{R} = \overline{r}.
\end{eqnarray}
The matrix $\Lambda_{\alpha\beta}$ is written in a similar form with the
$p_0\times p_0$ $R$-RSB Parisi matrices $\lambda_{ab}$ and $\rho_{ab}$.

At difference with the Parisi RSB scheme the block sizes $p_k$
are sent eventually to infinity 
in order, so that $p_k/p_{k-1}\to 0$, with the assumption that
\begin{equation}
\label{eq:DGO1}
  p_k\,(q_k - r_k) \to -\dot\Delta_{q_k}, \quad
  p_k\,(\lambda_k - \rho_k) \to -\dot\Delta_{\lambda_k}.
\end{equation}
as $p_k\to\infty$.
The limit $n\to 0$ is taken at the end.

As an example we consider
\begin{eqnarray}
\sum_{\alpha\beta}\Lambda_{\alpha\beta}Q_{\alpha\beta} &=& 
\nonumber\\
&\phantom{=}&
\hskip-60pt
n\sum_{t=0}^{R+1}p_t\bigl[
    (\lambda_t q_t - \rho_t r_t) - 
    (\lambda_{t-1} q_{t-1} - \rho_{t-1} r_{t-1})
                    \bigr]
\nonumber\\
&\phantom{=}&
\hskip-60pt
+ \frac{n^2}{p_0}\sum_{t=0}^{R+1}p_t
                (\rho_t r_t - \rho_{t-1} r_{t-1})
\end{eqnarray}
From Eq. (\ref{eq:DGO1}) we have 
$r_t = q_t + \dot\Delta_{q_t}/p_t + o(1/p_t)$
and a similar expression for $\rho_t$. As a consequence,
performing the ordered
limit $p_t\to\infty$  we have
\begin{equation}
\sum_{\alpha\beta}\Lambda_{\alpha\beta}~Q_{\alpha\beta} =
 n\Biggl[\overline{\lambda} - \lambda_R q_R - 
   \sum_{t=0}^{R} (\lambda_t\dot\Delta_{q_t} + q_t\dot\Delta_{\lambda_t})
   \Biggr]
\end{equation}

The evaluation of the trace is more involved. We shall give here the main 
steps. With the DGO form of $\Lambda_{\alpha\beta}$ we have
\begin{eqnarray}
\label{eq:DGO2}
\sum_{\alpha\beta}\Lambda_{\alpha\beta}\sigma_{\alpha}\sigma{\beta} &=&
(\overline{\lambda}-\lambda_R)\sum_{\alpha}\sigma_{\alpha}^2 
\nonumber\\
&\phantom{=}&
\hskip -60pt
+ \sum_{t=0}^{R+1}p_t\bigl[
    (\lambda_t q_t - \rho_t r_t) - 
    (\lambda_{t-1} q_{t-1} - \rho_{t-1} r_{t-1})\bigr] 
\nonumber\\
&\phantom{=}&
\hskip-20pt 
\times \sum_{i=1}^{p_0}\sum_{a_0\ldots a_{t-1}}
\left(\sum_{a_t\ldots a_R}\sigma^i_{a_0\ldots a_R}\right)^2
\\
&\phantom{=}&
\hskip -60pt 
+ \sum_{t=0}^{R}(\rho_t - \rho_{t-1})\sum_{a_0\ldots a_{t-1}}
\left(\sum_{i=1}^{p_0}\sum_{a_t\ldots a_R}\sigma^i_{a_0\ldots a_R}\right)^2
\nonumber
\end{eqnarray}
where the index $i=1,\ldots,p_0$ is relative to the primary 
blocks of size $p_0\times p_0$, while the index $a_k$ to the
sub-blocks of the Parisi RSB scheme.

By inserting this expression into the exponent of the exponential 
in the trace one ends up after a straightforward algebra with
\begin{eqnarray}
\label{eq:DGO3}
  \mbox{Tr}_{\sigma}
  \exp\left(\frac{1}{2}\sum_{\alpha\beta}\Lambda_{\alpha\beta}\,
   \sigma_{\alpha}\sigma_{\beta} +
            b\sum_{\alpha\beta}\sigma_\alpha\right) &=&
\nonumber\\
&\phantom{=}&
\hskip-150pt
\prod_{t=0}^{R}\left\{
         \prod_{a_0\ldots a_{t-1}}\left[
	     \int Dz_t\prod_i\int Dy_{i,t}
                                 \right]
	   \right\}
\nonumber\\
&\phantom{=}&
\hskip-120pt
\times
\prod_{a_0\ldots a_{R-1}}\prod_i \exp[p_R\,\phi_R(H_R)]
\end{eqnarray}
where $z_t \equiv z(a_0\ldots a_{t-1})$ and 
$y_{i,t}\equiv y_i(a_0\ldots a_{t-1})$ 
are the auxiliary Gaussian variables used to linearize the squares in 
(\ref{eq:DGO2}), and we have used the short-hand notation
\begin{equation}
Dz_t := \frac{dz_t}{\sqrt{2\pi}}\,e^{-z_t^2/2}, \quad
Dy_{i,t} := \sqrt{\frac{p_{t}}{2\pi}}dy_{i,t}\,e^{-p_{t}\,y_{i,t}^2/2}.
\end{equation}
The function $\phi_R(H)$ is defined as
\begin{eqnarray}
\exp[\phi_R(H)] &=& \mbox{Tr}_{\sigma}
          \exp\left[\frac{\overline{\lambda}-\lambda_R}{2}\sigma^2 + H\sigma
	      \right] 
\nonumber\\
&=& 
\sqrt{\frac{2\pi}{\lambda_R - \overline{\lambda}}}\,
    \exp\left[\frac{H^2}{2(\lambda_R - \overline{\lambda})}\right]
\end{eqnarray}
while $H_R$ is given by
\begin{equation}
\label{eq:br}
H_R = \sum_{t=0}^R\bigl[\sqrt{\Delta\lambda_t}\,z_t +
                 \sqrt{-\dot\Delta_{\lambda_t}}\,y_{i,t}
		 \bigr]
+ b
\end{equation}
with $\Delta\lambda_t = \lambda_t - \lambda_{t-1}$. In Eq. (\ref{eq:DGO3})
we used the fact that $H_R$ does not depend on $a_R$.

In the limit $p_R\gg 1$ the integral over $y_{i,R}$ can be 
evaluated at the saddle point
\begin{equation}
y_{i,R} = \sqrt{-\dot\Delta_{\lambda_R}}\, m_R
\end{equation}
where
\begin{equation}
\label{eq:DGOmR}
\bar m_{R} = \left.\frac{d}{dH}\phi_R(H)\right|_{H=H_R} = \phi_R'(H_R).
\end{equation}
Noticing that 
$\sum_{i} [\cdots]= O(n)$ and $n\ll 1$, we then have
\begin{eqnarray}
\int Dz_{R}\prod_{i=1}^{p_0} \int Dy_{i,R}\,
e^{p_R\,\phi_R(H_R)} &=& 
\\
\nonumber
&\phantom{=}&
\hskip -100pt
\prod_i \exp\left[p_{R-1}\,\phi_{R-1}(H_{R-1})\right]
\end{eqnarray}
where $H_{R-1}$ is given by Eq. (\ref{eq:br}) with the replacement 
$R\to\ R-1$, while

\begin{eqnarray}
\phi_{R-1}(H) &=& \int Dz_R\,
\Bigl[
 \frac{1}{2}\dot\Delta_{\lambda_R} \bar m_{R}^2
\nonumber\\
&\phantom{=}&
\hskip -10pt
 + \phi_R(
 \sqrt{\Delta\lambda_R}\,z_{R} - \dot\Delta_{\lambda_R} \bar m_{R} + H
         )
\Bigr].
\end{eqnarray}

The procedure can be repeated integrating over $y_{i,R-1}$, then over 
$y_{i,R-2}$ and so on. After having integrated out all $y_{i,t}$
we end up with
\begin{equation}
  \mbox{Tr}_{\sigma}
  \exp\left(\frac{1}{2}\sum_{\alpha\beta}\Lambda_{\alpha\beta}\,
   \sigma_{\alpha}\sigma_{\beta} +
            b\sum_{\alpha\beta}\sigma_\alpha\right) = 
\exp[{n\,\phi_{-1}(b)}]
\end{equation}
where
\begin{equation}
\phi_{-1}(b) = \int\prod_{t=0}^{R}\,Dz_t\,
   \Bigl[
 \frac{1}{2}\sum_{t=0}^R \dot\Delta_{\lambda_t}\, \bar m_t^2
 + \phi_R(H\{z\})
\Bigr]
\end{equation}
with
\begin{equation}
H(\{z\}) = \sum_{t=0}^{R}\Bigl[\sqrt{\Delta\lambda_t}\, z_t - 
                          \dot\Delta_{\lambda_t}\, \bar m_t\Bigr] + b
\end{equation}

and
\begin{eqnarray}
\bar m_t \equiv \bar m_t(z_0,\ldots,z_t) &=& 
\int Dz_{t+1}\, \bar m_{t+1}(z_0,\ldots,z_{t+1}) \nonumber\\
&=& 
\int \prod_{r=t+1}^{R} Dz_r\, \bar m_R(z_0,\ldots,z_R) \nonumber\\
\end{eqnarray}
with $\bar m_R(z_0,\ldots,z_R)$ defined by Eq. (\ref{eq:DGOmR}).

Collecting all terms, and redefining $\phi_R(H)$ as
\begin{equation}
\phi_R(H) = \frac{1}{2} \frac{H^2}{\lambda_R-\overline{\lambda}}
\end{equation}
to extract trivial factors, 
we obtain the Sompolinsky functional\cite{Somp81b,deDomGabOrl81}
for the spherical $2+p$ model
\begin{eqnarray}
\label{eq:fSomp}
-\beta f_{\rm S} &=& \frac{1}{2}\left[
  g(1) - q(q_R) - \overline{\lambda} + \lambda_R q_R
                        \right]
\\
&\phantom{=}&-\frac{1}{2}\left[
  \sum_{t=0}^{R}\Lambda(q_t)\,\dot\Delta_{q_t}
 -  \sum_{t=0}^{R} (\lambda_t\dot\Delta_{q_t} + q_t\dot\Delta_{\lambda_t})
\right] \nonumber\\
&\phantom{=}&
+ \int\prod_{t=0}^{R}\,Dz_t\,
   \left[
 \frac{1}{2}\sum_{t=0}^R \dot\Delta_{\lambda_t}\, \bar m_t^2
 + \phi_R(H(z))
\right]
\nonumber
\\
\nonumber
&\phantom{=}&
+ \frac{1}{2}\log\left(\frac{2\pi}{\lambda_R-\overline{\lambda}}\right).
\end{eqnarray}
The Sompolinsky solution follows from stationarity of $f_{\rm S}$ with
respect to variations of $\bar m_t$, $\dot\Delta_{\lambda_t}$,
$\Delta\lambda_t = \lambda_t-\lambda_{t-1}$ $\dot\Delta_{q_t}$, $q_t$,
and $\overline{\lambda}$ leading to, respectively,
\begin{equation}
\label{eq:Amr}
\bar m_r = \int Dz_{r+1}\, \bar m_{r+1}
\end{equation}
with $\bar m_R = \phi'_R(H\{z\})$,
\begin{equation}
q_r = \int\prod_{t=0}^R Dz_t\, \bar m_r^2
\end{equation}
\begin{equation}
\frac{1}{\lambda_R - \overline{\lambda}} 
 - \sum_{t=r}^R \dot\Delta_{q_t} =
\frac{1}{\sqrt{\Delta\lambda_r}} \int\prod_{t=0}^R Dz_t\, 
        \frac{\partial \bar m_r}{\partial z_r}
\end{equation}
\begin{equation}
\label{eq:Ala}
\lambda_r = \Lambda(q_r)
\end{equation}
\begin{equation}
\label{eq:Adot}
\dot\Delta_{\lambda_r} = \Lambda'(q_r)\,\dot\Delta_{q_r}
\end{equation}
and 
\begin{equation}
\label{eq:Aspc}
(\lambda_R -\overline{\lambda})^{-1} = 1 - q_R
\end{equation}
that is the spherical constraint.

By using the stationary equations (\ref{eq:Ala}), (\ref{eq:Adot}) and
(\ref{eq:Aspc})
to eliminate $\overline{\lambda}$, $\lambda_t$ and $\dot\Delta_{\lambda_t}$ 
from $f_{\rm S}$,
and changing the notation as
$\dot\Delta_{\lambda_t} \to \Delta'_t$,
$\dot\Delta_{q_t} \to \dot\Delta_t$ and
$\Delta\lambda_t \to \Delta_t = \Lambda(q_t) - \Lambda(q_{t-1})$
it is easy to see that the functional (\ref{eq:fSomp}) 
reduces to the Sompolinsky functional (\ref{eq:fsSomp}).

\section{$R\to\infty$ DGO theory}
\label{app:R_DGO}
To compare the DGO theory with the Parisi theory in the limit $R\to\infty$ 
we first eliminate the local magnetization $\bar m_r$ using the the stationary 
equation (\ref{eq:Amr}). For the 
the spherical $2+p$ spin glass model the equations can be easily solved 
obtaining
\begin{equation}
\bar m_r = \sum_{t=0}^{r} \frac{\sqrt{\Delta\lambda_t}}{F_t}\, z_t + \frac{b}{F_0}
\end{equation}
where
\begin{equation}
F_r = \lambda_R - \overline{\lambda} + \sum_{t=r}^R\dot\Delta_{\lambda_t}
\end{equation}
As a consequence
\begin{eqnarray}
\int\prod_{t=0}^RDz_t \sum_{t=0}^R\dot\Delta_{\lambda_t}\, \bar m_t^2 &=&
\nonumber\\
&\phantom{=}&
\hskip-50pt
      \sum_{r=0}^{R} \frac{\Delta\lambda_r}{F_r^2}
             \sum_{t=r}^{R}\dot\Delta_{\lambda_t}
+ \frac{b^2}{F_0^2}\sum_{t=0}^{R}\dot\Delta_{\lambda_t}
\end{eqnarray}
and
\begin{equation}
\int\prod_{t=0}^RDz_t\, H(\{z\})^2 = 
 (\lambda_R - \overline{\lambda})^{2}\left[
\sum_{t=0}^{R} \frac{\Delta\lambda_t}{F_t^2} + \frac{b^2}{F_0^2}
\right]
\end{equation}
Collecting all terms one finally has
\begin{eqnarray}
\label{eq:fDGO1}
&&-\beta f_{\rm DGO} = \frac{1}{2}\left[
  g(1) - g(q_R) - \overline{\lambda} + \lambda_R q_R
                        \right]
\\
&&\quad\quad-\frac{1}{2}\left[
   \sum_{t=0}^{R}\Lambda(q_t)\,\dot\Delta_{q_t}
  -   \sum_{t=0}^{R} (\lambda_t\dot\Delta_{q_t} + q_t\dot\Delta_{\lambda_t})
\right] \nonumber\\
\nonumber
&&\quad\quad+\frac{1}{2}\sum_{t=0}^R \frac{\lambda_t-\lambda_{t-1}}{F_t}
+\frac{1}{2}\frac{b^2}{F_0}
+\frac{1}{2}\log\left(\frac{2\pi}{\lambda_R-\overline{\lambda}}\right)
\nonumber
\end{eqnarray}
which is the more usual form of the DGO functional. Again the equations
for order parameters follow from stationarity of $f_{\rm DGO}$.
It can be checked that by eliminating the 
order parameters $\lambda_t$ and $\dot\Delta_{\lambda_t}$ and 
$\overline{\lambda}$ with the corresponding stationary equations
the DGO functional (\ref{eq:fDGO1}) reduces to 
the DGO functional (\ref{eq:fDGOr}) given in the main text.

In the limit $R\to\infty$, ad assuming that we are in a FRSB phase,
the DGO functional (\ref{eq:fDGO1}) of the spherical $2+p$ spin glass 
model becomes
\begin{eqnarray}
\label{eq:fDGO2}
-\beta f_{\rm DGO} &=& \frac{1}{2}\left[
  g(1) - g(q_1) - \overline{\lambda} + \lambda_1 q_1
                        \right]
\\
&\phantom{=}&-\frac{1}{2}
 \int_{0}^{1}dx\, \Lambda[q(x)]\,\dot\Delta_{q}(x)
\nonumber
\\
&\phantom{=}&+ \frac{1}{2} \int_{0}^{1}dx\, \bigl[\lambda(x)\,\dot\Delta_{q}(x) 
                 + q(x)\dot\Delta_{\lambda}(x)\bigr]
\nonumber
\\
&\phantom{=}&
+\frac{1}{2}\int_{0}^{1}dx \frac{\dot\lambda(x)}{F(x)}
+\frac{1}{2}\frac{\lambda(0) + b^2}{F(0)}
\nonumber
\\
&\phantom{=}&+\frac{1}{2}\log\left(\frac{2\pi}{\lambda_1-\overline{\lambda}}\right)
\nonumber
\end{eqnarray}
where $\dot\lambda(x) = d\lambda(x)/dx$, $\lambda_0 = \lambda(0)$ and
\begin{equation}
F(x) = \lambda_1 - \overline{\lambda} 
     +\int_{x}^{1}dx'\, \dot\Delta_{\lambda}(x').
\end{equation}

The expression (\ref{eq:fDGO2}) is specific of the spherical $2+p$ model, 
however it can be written in the more usual form for FRSB 
phase:\cite{Parisi80,deDomGabDup82,CriLeuPar02}
\begin{eqnarray}
-\beta f_{\rm DGO} &=& \frac{1}{2}\left[
  g(1) - g(q_1) - \overline{\lambda} + \lambda_1 q_1\right]
\nonumber\\
&\phantom{=}&
- \frac{1}{2}\int_{0}^{1}dx\, \Lambda[q(x)]\,\dot\Delta_{q}(x)
\nonumber\\
&\phantom{=}&
+\frac{1}{2}
   \int_{0}^{1}dx\, \bigl[\lambda(x)\,\dot\Delta_{q}(x) 
                 + q(x)\dot\Delta_{\lambda}(x)\bigr]
\nonumber\\
&\phantom{=}&
+\int_{-\infty}^{+\infty} \frac{dy}{\sqrt{2\pi\lambda(0)}}
\exp\left[-\frac{(y-b)^2}{2\lambda(0)}\right]\,\phi(0,y)
\nonumber\\
&\phantom{=}&
+ \frac{1}{2}\log\left(\frac{2\pi}{\lambda_1-\overline{\lambda}}\right)
\end{eqnarray}
where
\begin{equation}
\phi(x,y) = \frac{1}{2}\left[
        \frac{y^2}{F(x)} + \int_{x}^{1}dx' \frac{\dot\lambda(x')}{F(x')}
\right]
\end{equation}
is solution of the Parisi anti-parabolic differential equation
\begin{equation}
\dot\phi(x,y) = - \frac{\dot\lambda(x)}{2}\,\phi''(x,y) 
                + \frac{\dot\Delta_{\lambda}(x)}{2}\,\phi'(x,y)^2
\end{equation}
with the boundary condition
\begin{equation}
\phi(1,y) = \frac{1}{2}\frac{y^2}{\lambda_1 - \overline{\lambda}}.
\end{equation}
As usual a ``dot'' in the Parisi equation 
denotes the derivative with respect to $x$ while a
``prime'' the derivative with respect to $y$.

The Parisi solution is recovered by setting
$\dot\Delta_{\lambda} = -x\,\dot\lambda(x)$, 
$\dot\Delta_{q} = -x\,\dot q(x)$, see e.g. Ref [\onlinecite{CriLeu06}].


\section{Stability of the DGO-Sommers solution}
\label{app:StabSom}
In the DGO$_{R=0}$ Ansatz the free energy fluctuations in the replica space,
cfr. Eq. (\ref{eq:stabil}), becomes
\begin{eqnarray}
&&\delta^2\left[-\beta f(r,q,m)\right]
\\
\nonumber
&&\quad=-\frac{1}{n}\sum_{ab}\left\{ \Lambda'(r)+\hat\epsilon_{ab}\left[
\Lambda'(q)-\Lambda'(r)
\right]
\right\}\left(\delta q_{ab}\right)^2
\\
\nonumber
&&\quad \quad+\frac{A^2}{n}\sum_{ab}\left(\delta q_{ab}\right)^2 
+ \frac{B^2}{n}\mbox{Tr} \left(\hat{\bm{\epsilon}}~ \bm{\delta q}\right)^2
\\
\nonumber
&&\quad \quad+C^2\left(\sum_{ab}\delta q_{ab}\right)^2
+2AB~\mbox{Tr}~ \bm{\delta q} ~\hat{\bm \epsilon} ~\bm{\delta  q}
\\
\nonumber
&&\quad \quad
+2AC\sum_{ab}\left(\bm{\delta  q}~ \bm{\delta  q} \right)_{ab}
+2BC\sum_{ab}\left(\bm{\delta  q}~ \hat{\bm \epsilon}~ \bm{\delta  q}\right)_{ab}
\end{eqnarray}

with 
\begin{eqnarray}
A&=&=\frac{1}{1-q}
\\
B&=&-\frac{q-r}{(1-q)\chi_1}
\\
C&=&-\frac{r}{\chi_1}
\\
\chi_1&=&1-q-\dot \Delta
\end{eqnarray}
where $\dot \Delta = -p_0 (q-r)$ by definition.
The eigenvalue equation is
\begin{eqnarray}
&&\left[A^2-\Lambda'(r)\right]\delta q_{ab}
-\left[\Lambda'(q)-\Lambdaì(r)\right]\hat\epsilon_{ab}~\delta q_{ab}
\label{app:eigenvalues}
\\
\nonumber
&&\ \ +B^2\left(\hat{\bm \epsilon} ~\bm {\delta  q} ~\hat{\bm \epsilon} \right)_{ab}
+C^2\left(\sum_{cd}\delta q_{cd}\right)\delta q_{ab}
\\
\nonumber
&&\ \ +AB\left[\left(\hat{\bm \epsilon} ~\bm{\delta  q}\right)_{ab}+
\left(\bm{\delta  q}
 ~\hat{\bm \epsilon}\right)_{ab}\right]
+AC\sum_c\left(\delta q_{ac}+\delta q_{bc}\right)
\\
\nonumber
&&\ \ +BC\sum_c\left[\left(\hat{\bm \epsilon} ~\bm{\delta  q}\right)_{ac}+
\left(\hat{\bm \epsilon}~\bm{\delta  q} \right)_{bc}\right]
=\lambda ~\delta q_{ab}
\end{eqnarray}
The above equation  is valid for $a\neq b$. The diagonal elements
 $\delta q_{aa}$ are all zero because of the spherical constraint.
  In the present Ansatz we have $n/p_0$ blocks each containing $p_0$
 elements. The diagonal blocks contain $q$-elements, whereas the
 off-diagonal ones contain $r$-elements. $q$ is the overlap value of
 replicas belonging to the same  cluster, $r$ the overlap between
 replicas of different clusters.  The different eigenvalues,
 solutions of Eq. (\ref{app:eigenvalues}), can be grouped in three
 different sets each one corresponding to a given  subspace of the
 replica space. One subspace involves fluctuations of the overlaps
 of one replica with other $p_0$ replicas (both belonging to the same
 cluster and different clusters). Another one involves fluctuations
 of the overlaps of groups of $p_0$ replicas with other $p_0$
 replicas. The third one consists in the eigenvalues determining the
 stability of the fluctuations between clusters as a whole (roughly
 speaking). We look in detail at the eigenvalues and at their
 behavior as $n\to 0$. 
\subsection{Fluctuations of the  overlaps of one  replica with 
${\bm p_0}$ other replicas}

The first  subspace is
determined by the condition
\begin{equation}
\left(\hat{\bm \epsilon}~\bm{\delta q}\right)_{ab}=0
\quad \quad \forall~ a,b
\end{equation}
Two eigenvalues are associated to this subspace. 
One corresponds to fluctuations
of the overlap between replicas in two different clusters (off-diagonal 
elements), for which all diagonal blocks are zero:
\begin{equation}
\label{app:1m_subspace}
\hat\epsilon_{ab}~\delta q_{ab}=0
\quad\quad \forall ~a,b
\end{equation}
The eigenvalue and its degeneracy are: 
\begin{eqnarray}
\Lambda^{(1)}_0&=&-\Lambda'(r)+A^2
\label{app:lambda10}
\\
n^{(1)}_0&=&\frac{n(n-p_0)(p_0-1)^2}{2p_0^2}
\nonumber
\end{eqnarray}

The other one controls fluctuations of the $q$-overlaps, i.e.
the off-diagonal blocks are zero: 
\begin{equation}
(1-\hat\epsilon_{ab})\delta q_{ab}=0
\quad \forall ~ a,b
\end{equation}
Its expression and its degeneracy are: 
\begin{eqnarray}
\Lambda^{(1)}_1&=&-\Lambda'(q)+A^2
\label{app:lambda11}
\\
n^{(1)}_1&=&\frac{n(p_0-3)}{2}
\nonumber
\end{eqnarray}

\subsection{Fluctuations of the overlaps of ${\bm p_0}$ replicas with other 
${\bm p_0}$ replicas. }

We now look at the fluctuations in the subspace
\begin{equation}
\label{app:mm_subspace}
\left(\hat{\bm \epsilon}~\bm{\delta q}~\hat{\bm \epsilon}\right)_{ab}=0
\quad \quad \forall ~a,b
\end{equation}
with  $\left(\hat{\bm \epsilon}~\bm {\delta q }\right)_{ab}\neq 0$, [Eq.
(\ref{app:1m_subspace}) not satisfied].

The first eigenvalue can be addressed as the one  related to 
 fluctuations between different clusters as a whole, that is the subspace
given by the further condition
\begin{equation}
\hat\epsilon_{ab}\left(\hat{\bm \epsilon}~\bm{\delta q}\right)_{ab}=0
\quad \quad \forall a,b
\label{app:sub2}
\end{equation}

Eigenvalue and degeneracy are:
\begin{eqnarray}
\Lambda^{(2)}_0&=&-\Lambda'(r)+A^2+p_0AB
\label{app:lambda20}
\\
n^{(2)}_0&=&\frac{n(n-p_0)(p_0-1)}{p_0^2}
\nonumber
\end{eqnarray}

The second eigenvalue deals with the subspace orthogonal
to Eq. (\ref{app:sub2}), i.e. with fluctuations between replicas in the 
same cluster:
\begin{equation}
(1-\hat\epsilon_{ab})\left(\hat{\bm \epsilon}~\bm{\delta q}\right)_{ab}=0
\quad \quad \forall a,b
\end{equation}
Its form and degeneracy are:
\begin{eqnarray}
\Lambda^{(2)}_1&=&-\Lambda'(q)+A^2 +(p_0-2)A(B+C)
\label{app:lambda21}
\\
\nonumber
n^{(2)}_1&=&\frac{n(p_0-1)}{p_0}
\end{eqnarray}


\subsection{Fluctuations of the overlap of one cluster with other clusters}
Here we consider the clusters as single elements and the relative
 fluctuations. The subspace we look at is orthogonal to the first two 
subspaces and in order to express the condition defining it we introduce
the 
cluster matrix
\begin{equation}
\label{app:cc_subspace}
 \mathbf{C}_{\alpha\beta}= \left(\hat{\bm \epsilon}~\bm{\delta q}\right)_{ab}
\ \ \ \quad \mbox{ with } a\in\alpha,  \ \ b\in \beta
\end{equation}
$\alpha,\beta$ are cluster indexes.
In terms of this matrix one identifies a first sub-sub-space 
associated with purely off-diagonal fluctuations (i.e. between different
clusters):
\begin{equation}
\bm C_{\alpha\alpha}=0 \ \ \quad\quad 
\sum_\beta \bm C_{\alpha\beta}=0\quad \forall \alpha
\end{equation}

The eigenvalue and its degeneracy are
\begin{eqnarray}
\Lambda^{(3)}_{0}&=&-\Lambda'(r)+(A+p_0B)^2
\label{app:lambda30}
\\
n^{(3)}_0&=&\frac{n(n-3p_0)}{2p_0^2}
\end{eqnarray}

There are, then, other two subspaces (for finite $n$), whose physical
meaning is less clear since mixed fluctuations are involved.

One subsubspace is determined by the eigenvectors for which
\begin{equation}
\label{app:sub312}
\sum_\alpha\bm C_{\alpha\alpha}=0 \ \ \quad\quad 
\sum_{\alpha\neq \beta} \bm C_{\alpha\beta} =0
\end{equation}
Defining 
\begin{eqnarray}
&&U=-\Lambda'(r)-\Lambda'(q)+2(A+p_0B)^2
\\
\nonumber
&&\hspace*{1cm }-B ~(2A+p_0B)+W+Z
\\
&&V=-WZ+\Bigl[
-\Lambda'(r)+(A+p_0B)^2+W
\Bigr]
\\
\nonumber
&&\hspace*{.1cm }\times \Bigl[-\Lambda'(q)+(A+p_0B)^2
-B~(2A+p_0B)+Z
\Bigr]
\nonumber 
\\
&&W=(n-2p_0)~C~(A+p_0B)
\\
&&Z=2(p_0-1)~C~(A+p_0B)
\end{eqnarray}
the  two eigenvalues are 
\begin{eqnarray}
\Lambda^{(3)}_{1,2}&=&\frac{U}{2}\Biggl[1\pm\sqrt{1-4 \frac{U}{V}}\Biggr]
\label{app:lambda312}
\\
n^{(3)}_{1,2}&=&\frac{n-p_0}{p_0}
\end{eqnarray}

The last subspace is set by the eigenvectors orthogonal to 
Eq. (\ref{app:sub312}):
\begin{equation}
\sum_\alpha\bm C_{\alpha\alpha} \neq 0 \ \ \quad\quad 
\sum_{\alpha\neq \beta} \bm C_{\alpha\beta} \neq 0
\end{equation}
Also in this case there are two different eigenvalues, whose expression is
identical to Eq (\ref{app:lambda312}) provided that 
$U=2(n-p_0)C(A+p_0B)+n(n-p_0)(C^2+b^2)$. Their degeneracy is $1$.

\section{Dynamical Solution for the $2+p$ spherical model}
\label{app:Dyn2p}
In this Appendix we show that the CHS dynamical solution 
of the spherical $2+p$ spin glass model requires marginality of the
dynamics in the intermediate timescales. To keep the notation simple
and to refer to a physically well known system,
we shall consider the case of two time scales, appropriate for
the 1RSB-type phase. With minor changes the derivation can be 
generalized to any number of time scales.

By inserting the forms (94)-(95) for the correlation and response function
into the Dyson equation (\ref{eq:Dyson}) and separating out the 
short time behavior $\omega\gg \epsilon$ as $\epsilon\to 0$ and long time 
behavior $\omega = \epsilon\,\Omega$ as $\epsilon\to 0$, 
one obtains the following equations of motion 
for $G_1(\omega)$ and $G_0(\Omega)$:
\begin{equation}
\label{eq:G1}
\left(r - \frac{i\omega}{\Gamma_0}\right)\,G_1(\omega) -
 \Sigma_1(\omega)\,G_1(\omega) = 1
\end{equation}
\begin{eqnarray}
\left(\overline{r} + \Lambda(q_1) - \epsilon\frac{i\Omega}{\Gamma_0}\right)\,
G_0(\Omega) &-& (1-q_1)\,\Sigma_0(\Omega) \nonumber\\
&\phantom{-}& \hskip-1cm
- \Sigma_0(\Omega)\,G_0(\Omega) = 0
\label{eq:G0}
\end{eqnarray}
where $\Sigma_1(\omega)$ and $\Sigma_0(\Omega)$ are the short and long time 
part of the self-energy $\Sigma(\omega)$, and
\begin{equation}
\label{eq:overr}
\overline{r} \equiv r - \Lambda(1) = -\Lambda(q_1) + \frac{1}{1-q_1}
\end{equation}
to ensure the correct static limit $\omega\to 0$ of Eq. (\ref{eq:G1}).
The static limit $\Omega\to 0$ of Eq. (\ref{eq:G0}) gives the equation for
$q_0$.
\begin{equation}
\Lambda(q_1) - \Lambda(q_0) = 
(\overline{r} + \Lambda(q_1))\,\frac{q_1-q_0}{1-q_1 + m(q_1-q_0)}.
\end{equation}
The parameter $\overline{r}$ can be 
 eliminated from these equations with the help
of the spherical constraint
\begin{eqnarray}
\int_{-\infty}^{+\infty}\frac{d\omega}{2\pi}C(\omega) &=& 
    2 (1-q_1) - \overline{r}(1-q_1)^2 
\nonumber\\
&\phantom{=}& + 2 m (1-q_1)(q_1-q_0)\Lambda(q_1) 
\nonumber\\
&\phantom{=}& - m\left(\overline{r} + (1-m)\Lambda(q_1)\right) (q_1-q_0)^2
\nonumber \\
&=& 1
\end{eqnarray}
One, then, recovers Eqs. (\ref{eq:q1RSB1},\ref{eq:q0RSB1}) of the main text.

The equations for the correlation functions $C_1$ and $C_0$ are obtained 
from Eqs. (\ref{eq:G1}), (\ref{eq:G0}) by using the relations:
\begin{eqnarray}
G_1(\omega) &=& (1-q_1) + i\omega\int_{0}^{\infty}dt\,e^{i\omega t} C_1(t)
\nonumber\\
&=& (1-q_1) + i\omega\hat{C}_1(\omega)
\end{eqnarray}

\begin{eqnarray}
G_0(\omega) &=& m (1-q_1) 
+ m ~i\omega\int_{0}^{\infty}dt\,e^{i\omega t} \left(C_0(t) - q_0\right)
\nonumber\\
&=& m(q_1-q_0) + m~ i\omega\hat{C}_0(\omega)
\end{eqnarray}
and
\begin{eqnarray}
\Sigma_1(\omega) &=& \Lambda(1)-\Lambda(q_1) \nonumber\\
&\phantom{=}&
+ i\omega\int_{0}^{\infty}dt\,e^{i\omega t} 
      \left(\Lambda[C_1(t)+q_1] - \Lambda(q_1)\right)
\nonumber\\
&=& \Lambda(1)-\Lambda(q_1) + i\omega\hat{\Lambda}_1(\omega)
\end{eqnarray}

\begin{eqnarray}
\Sigma_0(\omega) &=& m\left(\Lambda(q_1)-\Lambda(q_0)\right) \nonumber\\
&\phantom{=}&
+ m~ i\omega\int_{0}^{\infty}dt\,e^{i\omega t} 
      \left(\Lambda[C_0(t)] - \Lambda(q_0)\right)
\nonumber\\
&=& m\left(\Lambda(1)-\Lambda(q_1)\right) + m ~i\omega\hat{\Lambda}_0(\omega)
\end{eqnarray}
that follows from FDT. A simple algebra leads to the equations
\begin{eqnarray}
\left(\overline{r} + \Lambda(q_1) -\frac{i\omega}{\Gamma_0}\right) 
\hat{C}_1(\omega) 
    &-& \hat{\Lambda}_1(\omega)G_1(\omega) \nonumber\\
&-&\frac{1}{\Gamma_0}(1-q_1) = 0
\end{eqnarray}
and 
\begin{eqnarray}
&-&\frac{\epsilon}{\Gamma_0}\left[q_1-q_0 + i\Omega \hat{C}_0(\Omega) \right]
\nonumber\\
&\phantom{-}&\hskip1cm
+\bigl[\overline{r}+\Lambda(q_1) - m(\Lambda(q_1) - \Lambda(q_0))\bigr]
\hat{C}_0(\Omega)
\nonumber\\
&\phantom{-}&\hskip1cm
    -(1-q_1)\hat{\Lambda}_0(\Omega)
    - \hat{\Lambda}_0(\Omega)G_0(\Omega)
= 0
\end{eqnarray}

To study the stability of the static limits it is useful to rewrite these 
equation in the time space in the following equivalent form:
\begin{widetext}
\begin{eqnarray}
\label{eq:C1sp}
\Gamma_0^{-1}\partial_t C_1(t) 
 + \bigl[\overline{r}_1[C_1(t)+q_1] - \overline{r}\bigr](1-q_1 - C_1(t))
+\int_{0}^{t}dt' \bigl[\Lambda[C_1(t-t')+q_1] - \Lambda[C_1(t)+q_1]\bigr]
                   \partial_{t'}C_1(t') = 0
\end{eqnarray}
\begin{eqnarray}
\label{eq:C0sp}
\epsilon\Gamma_0^{-1}\partial_t C_0(t) 
 + \bigl[\overline{r}_0[C_0(t),m] - \overline{r}\bigr]
[1-q_1+m(q_1-C_0(t))]
+ m \int_{0}^{t}dt' \bigl[\Lambda[C_0(t-t')] - \Lambda[C_0(t)]\bigr]
                   \partial_{t'}C_0(t') = \delta
\end{eqnarray}
\end{widetext}

where $\delta = \lim_{t\to 0^+}\epsilon\Gamma_0^{-1}\partial_t C_0(t)$ and
$\overline{r}_1(q)$ and $\overline{r}_0(q,m)$ are the 
functions
\begin{equation}
\overline{r}_1(q) = -\Lambda(q) + \frac{1}{1-q}
\end{equation}
\begin{eqnarray}
\overline{r}_0(q,m) &=& 
    \overline{r}_1(q) \nonumber\\
&\phantom{=}&
     -\frac{(1-m)(q_1-q)^2}
                 {(1-q_1)(1-q)[1-q_1+m(q_1-q)]}.
\nonumber\\
\end{eqnarray}

\begin{figure}[t!]
\includegraphics[scale=1.0]{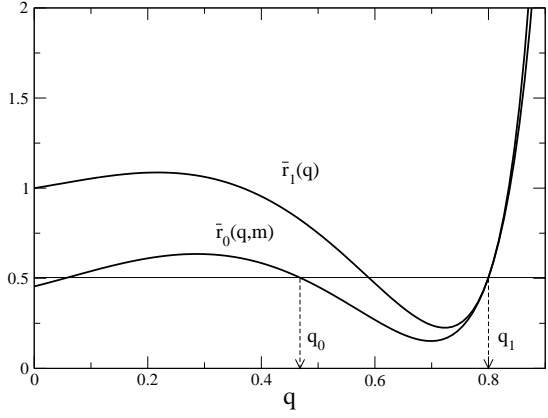}
\caption{Schematic form of $\overline{r}_1(q)$ and $\overline{r}_0(q,m)$ in the
         1RSB phase. The horizontal line shows the value of $\overline{r}$.
         In the plot the slope of $\overline{r}_1(q)$ at $q_1$ is strictly positive, implying
         that the slope of $\overline{r}_0(q,m)$ at $q_0$ (the largest solution 
         of the equation $\overline{r}_0(q,m) = \overline{r}$ below $q_1$) cannot be
         positive.
        }
\label{fig:nogood}
\end{figure}

In terms of these equations the physical values of $q_1,q_0$ ($0\leq
q_0 \leq q_1\leq 1$) yielding the {\it plateau} are the largest
solution of the equation\cite{CriHorSom93}
\begin{equation}
\overline{r}_1(q_1) = \overline{r}_0(q_0,m) = \overline{r}
\end{equation}
with $\overline{r}$ fixed by the spherical constraint.

\begin{figure}[t!]
\includegraphics[scale=1.0]{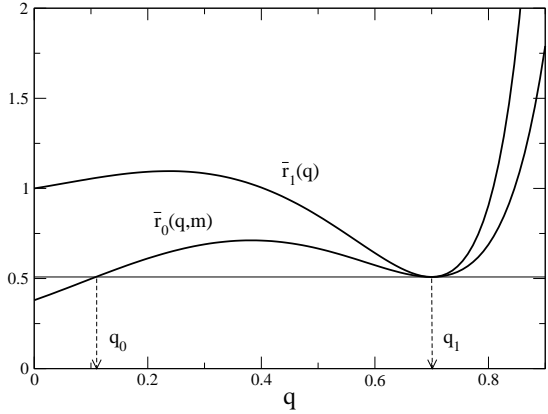}
\caption{Schematic form of $\overline{r}_1(q)$ and $\overline{r}_0(q,m)$ in the
         1RSB phase. The horizontal line shows the value of $\overline{r}$.
         Here the slope of $\overline{r}_1(q)$ at $q_1$ is zero, implying
         a positive slope of the function $\overline{r}_0(q,m)$ at $q_0$.        }
\label{fig:good}
\end{figure}

Expanding equations (\ref{eq:C1sp}), (\ref{eq:C0sp}) near the plateau
to the first order in the deviation one obtains the dynamic stability 
conditions
\begin{equation}
\left.\frac{\partial}{\partial q}\overline{r}_1(q)\right|_{q=q_1}\geq 0
\end{equation}
\begin{equation}
\left.\frac{\partial}{\partial q}\overline{r}_0(q,m)\right|_{q=q_0}\geq 0
\label{app:r0_cond}
\end{equation}
It is easy to check that these coincide with dynamical stability conditions 
(\ref{eq:stb1CHS}), (\ref{eq:stb0CHS}) given in the main 
text.

In the 1RSB phase $\overline{r}_1(q)$ and $\overline{r}_0(q,m)$ have the
shape depicted in Figs. \ref{fig:nogood} and \ref{fig:good}, while
$\overline{r}<1$.\cite{CriHorSom93} A simple analysis of these figures 
shows that in order to 
satisfy Eq. (\ref{app:r0_cond})
have $(\partial/\partial q)\overline{r}_0(q_0,m) > 0$
it is necessary that 
\begin{equation}
\left.\frac{\partial}{\partial q}\overline{r}_1(q)\right|_{q=q_1}= 0
\end{equation}
i.e., the solution at shorter time scales must be marginally stable.



\begin{thebibliography}{99}
\bibitem{Note-Luca}
 This conclusion can be extended to systems showing a ``one
+ full'' RSB phase, i.e., the phase described by an order parameter
function $q(x)$ composed by a continuous part plus a discontinuous
jump.  In the discussion of the infinite replica symmetry breaking
limit we shall always assume that the system is in the FRSB phase. The
extension to the 1-FRSB phase is straightforward and does not add much
to the discussion making, on the contrary, the formulae more heavy.
\bibitem{Note:beta}
  To simplify the notation we absorb the temperature $T$ into the definition
  of the correlation function.
\bibitem{Note:0} 
  In Eq. (\ref{eq:Dyson}) the subscript $0$
  meant the bare propagator, whereas here it stays for the long time 
  relaxation part  of the correlation and response functions.
\bibitem{Note-stab} We assume that the relaxation to equilibrium on the 
                    longest time-scale is non-singular, i.e., exponential.
                    If this is not the case, e.g. a power law decay, then
                    $\Gamma(\omega_0)=0$. This, however, does not invalidate 
                    the conclusions about the intermediate time sectors 
		    since dynamical stability requires that 
                    $\Gamma(\omega_0)$ be non-negative.
\bibitem{Note:MarStates}
  States other than genuine minima
  actually exist: the ``marginal states''.\cite{MueLeuCri06} 
  That is, stationary points of the free energy landscape
  whose Hessian spectrum of eigenvalues is always positive but for an
  isolated, null, eigenvalue. This states can dominate at higher free
  energy values, above the threshold ones, displaying their own 
  complexity.\cite{CriLeuRiz05}
  Their topology, however, is such
  that the relaxation dynamics - though slowed down by the difficulty of
  finding the unique way out - is not stuck forever and the dynamical
  arrest only occurs when the threshold states (minima) are reached.  
  In the present work we are considering exclusively the contribution to the
  dynamics of the genuine minima.
\end{thebibliography}
\end{document}